\documentclass[prd, preprint, showpacs,amsmath,amssymb,superscriptaddress,floatfix,nofootinbib,11pt]{revtex4}
\usepackage{mathrsfs,bm}
\usepackage{longtable,lscape}
\usepackage{txfonts}
\usepackage{amssymb}
\usepackage{rotating}
\usepackage{indentfirst}
\usepackage{graphicx,,booktabs}
\usepackage{multirow}
\usepackage{slashed}
\usepackage{overpic}
\usepackage{color}
\usepackage{amssymb}
\usepackage{hyperref}

\begin{document}
\title{Magnetic moments of the spin-1/2 singly charmed baryons in covariant baryon chiral perturbation theory}

\author{Rui-Xiang Shi}
\affiliation{School of Physics and
Nuclear Energy Engineering \& International Research Center for Nuclei and Particles in the Cosmos \&
Beijing Key Laboratory of Advanced Nuclear Materials and Physics,  Beihang University, Beijing 100191, China}

\author{Yang Xiao}
\affiliation{School of Physics and
Nuclear Energy Engineering \& International Research Center for Nuclei and Particles in the Cosmos \&
Beijing Key Laboratory of Advanced Nuclear Materials and Physics,  Beihang University, Beijing 100191, China}

\author{Li-Sheng Geng}
\email[E-mail: ]{lisheng.geng@buaa.edu.cn} \affiliation{School of
Physics and Nuclear Energy Engineering \& International Research
Center for Nuclei and Particles in the Cosmos \& Beijing Key
Laboratory of Advanced Nuclear Materials and Physics,  Beihang
University, Beijing 100191, China}
\begin{abstract}
Recent experimental advances have reignited theoretical interests in  heavy-flavor hadrons. In this work, we study the magnetic moments
of the spin-1/2 singly charmed baryons up to the next-to-leading order in covariant baryon chiral perturbation theory with the extended-on-mass-shell renormalization scheme.
The pertinent  low energy constants $g_{1-4}$ are fixed with the help of the quark model and the heavy quark spin flavor symmetry, while the remaining $d_2$, $d_3$, $d_5$ and $d_6$ are determined by fitting to the
lattice QCD pion-mass dependent data. We study the magnetic moments as a function of $m_\pi^2$ and compare our results with those
obtained in the heavy baryon chiral perturbation theory. We find that the loop corrections induced by the anti-triplet states are dominated by the baryon pole diagram. In addition,
we predict the magnetic moments of the spin-1/2 singly charmed baryons and compare them with those of other approaches.

\end{abstract}


\maketitle
\section{INTRODUCTION}
In the last two decades, tremendous progress has been made in our understanding of heavy-flavor
hadrons, thanks to the experimental discoveries by collaborations such as LHCb, BELLE, and BESIII and the related theoretical studies.
 In the charmed baryon sector,  24  singly charmed baryons and two doubly charmed baryons are listed in the current version of the review of particle physics~\cite{Tanabashi:2018oca}. Among
them, the newest members include the $\Lambda_c(2860)$~\cite{Aaij:2017vbw}, the five $\Omega_c$ states~\cite{Aaij:2017nav}, and the $\Xi_{cc}^{++}$~\cite{Aaij:2017ueg}. Inspired by these and other experimental discoveries, there are extensive theoretical and lattice QCD  studies on their nature and their decay and production mechanisms (see, e.g., Refs.~\cite{Chen:2016qju,Chen:2016spr,Guo:2017jvc,  
  Esposito:2016noz,Lebed:2016hpi,Eichmann:2016yit,Olsen:2017bmm,Ali:2017jda} references cited therein).

The magnetic moment of a baryon plays an extremely important role in understanding its internal structure. Historically, the experimental measurement of the magnetic moments of the proton and the neutron  revealed that they are not point-like particles. The subsequent studies helped the establishment of the quark model as well the theory of the strong interaction, Quantum Chromo Dynamics.  Unlike those of the ground-state baryons, the magnetic moments of the spin-1/2 singly charmed baryons have not been measured experimentally. Nevertheless,  they have been studied in a variety of phenomenological models~\cite{Barik:1984tq,JuliaDiaz:2004vh,Kumar:2005ei,Faessler:2006ft,Patel:2007gx,Sharma:2010vv,Bernotas:2012nz} as well as QCD sum rules~\cite{Zhu:1997as}. Lately, they have also been studied in the mean-field approach~\cite{Yang:2018uoj}, the self-consistent SU(3) chiral quark-soliton model~\cite{Kim:2018nqf}, the heavy baryon chiral perturbation theory (HB ChPT)~\cite{Wang:2018xoc}, and
lattice QCD simulations~\cite{Can:2013tna,Can:2015exa,Bahtiyar:2016dom,Bahtiyar:2015sga}. In Ref.~\cite{Wang:2018xoc}, the low energy constants (LECs) are determined by the quark model and the heavy quark spin flavor symmetry and by fitting to the lattice QCD data extrapolated to the physical point. In this work, we will study the magnetic moments of the spin-1/2 singly charmed baryons up to the next-to-leading order (NLO) in covariant baryon chiral perturbation theory (BChPT) with the extended-on-mass shell (EOMS) renormalization scheme. The unknown LECs will be
 determined by the quark model and the heavy quark spin flavor symmetry and by directly fitting to the lattice QCD data at unphysical pion masses~\cite{Bahtiyar:2016dom,Can:2013tna,Can:2015exa}. One notes that many previous studies, such as Refs.~\cite{Ren:2012aj,Xiao:2018rvd}, have shown that the EOMS BChPT can provide a better description of the lattice QCD quark-mass dependent data  than its non-relativistic counterpart.

Chiral perturbation theory (ChPT)~\cite{Weinberg:1978kz}, as a low-energy effective field theory of QCD,  is an appropriate framework to study the magnetic moments of hadrons, particularly, their light quark mass dependence. It
 provides a systematic expansion of physical observables in powers of $(p/\Lambda_\chi)^{n_\chi}$, where $p$ is a small momentum and  $\Lambda_\chi$ is the chiral symmetry breaking scale. However, its application to the one-baryon sector encountered a  difficulty, i.e., a systematic power counting (PC) is lost due to the large
non-vanishing baryon mass $m_0$ in the chiral limit. Over the years, three approaches were proposed to overcome this issue, i.e., the HB~\cite{Jenkins:1990jv,Bernard:1995dp}, the infrared (IR)~\cite{Becher:1999he}, and the EOMS~\cite{Fuchs:2003qc} schemes. The IR and the EOMS schemes are the relativistic formulations of BChPT. A brief summary and comparison of the three different approaches can be found in Ref.~\cite{Geng:2013xn}.

The EOMS scheme is different from the HBChPT, because it  retains a series of higher-order terms within the covariant power counting (PC) rule when removing the power-counting-breaking (PCB) terms. In recent years,
 many physical observables have been successfully studied in this scheme such as the magnetic moments~\cite{Xiao:2018rvd,Geng:2009ys,Geng:2008mf,Geng:2009hh,Liu:2018euh,Blin:2018pmj}, the masses and sigma terms~\cite{Ren:2012aj,Ren:2013oaa,Sun:2016wzh,Yao:2018ifh} of the octet, decuplet and spin-1/2 doubly heavy baryons, the hyperon vector couplings~\cite{Geng:2009ik,Geng:2014efa}, the axial vector charges~\cite{Ledwig:2014rfa}, the pion nucleon~\cite{Alarcon:2012kn,Chen:2012nx} and kaon-nucleon scattering~\cite{Lu:2018zof}. Thus, inspired by these studies, we would like to study the magnetic moments of the spin-1/2 singly charmed baryons in the EOMS scheme.

This work is organized as follows. In Sec.~II, we provide the effective Lagrangians and calculate the relevant Feynman diagrams up to ${\cal O}(p^3)$.
Results and discussions are given in Sec.~III, followed by a short summary in Sec. IV.
\section{Theoretical formalism}
The magnetic moments of singly charmed baryons are defined via the matrix elements of the electromagnetic current $J_\mu$ as follows:
\begin{eqnarray*}
\langle\psi(p_f)|J_\mu|\psi(p_i)\rangle=\bar{u}(p_f)\left[\gamma_\mu F_1^B(q^2)+\frac{i\sigma_{\mu\nu}q^\nu}{2m_B}F_2^B(q^2)\right]u(p_i),
\end{eqnarray*}
where $\bar{u}(p_f)$ and $u(p_i)$ are the Dirac spinors, $m_B$ is the singly charmed baryon mass, and $F_1^B(q^2)$ and $F_2^B(q^2)$ denote the Dirac and Pauli form factors, respectively. The
four-momentum transfer is defined as $q=p_i-p_f$. At $q^2=0$, $F_2^B(0)$ is the so-called anomalous magnetic moment, $\kappa_B$, and the magnetic moment is $\mu_B=\kappa_B+Q_B$, where $Q_B$ is the charge of the singly charmed baryon.
\begin{figure}[h!]
  \centering
  \includegraphics[width=4cm]{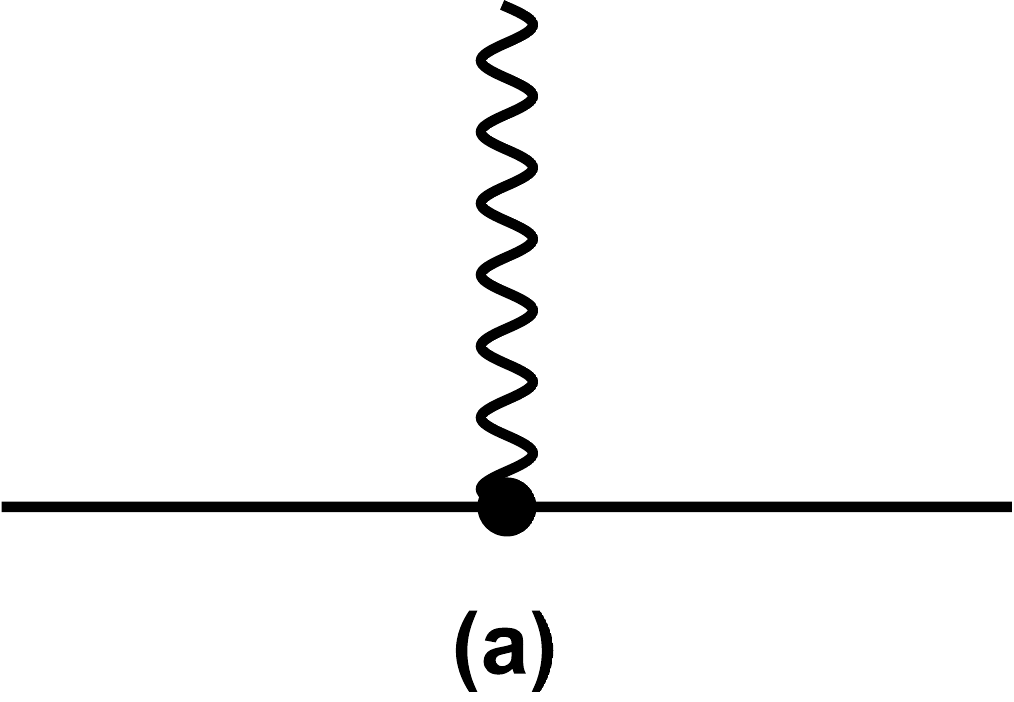}\\
  \includegraphics[width=4cm]{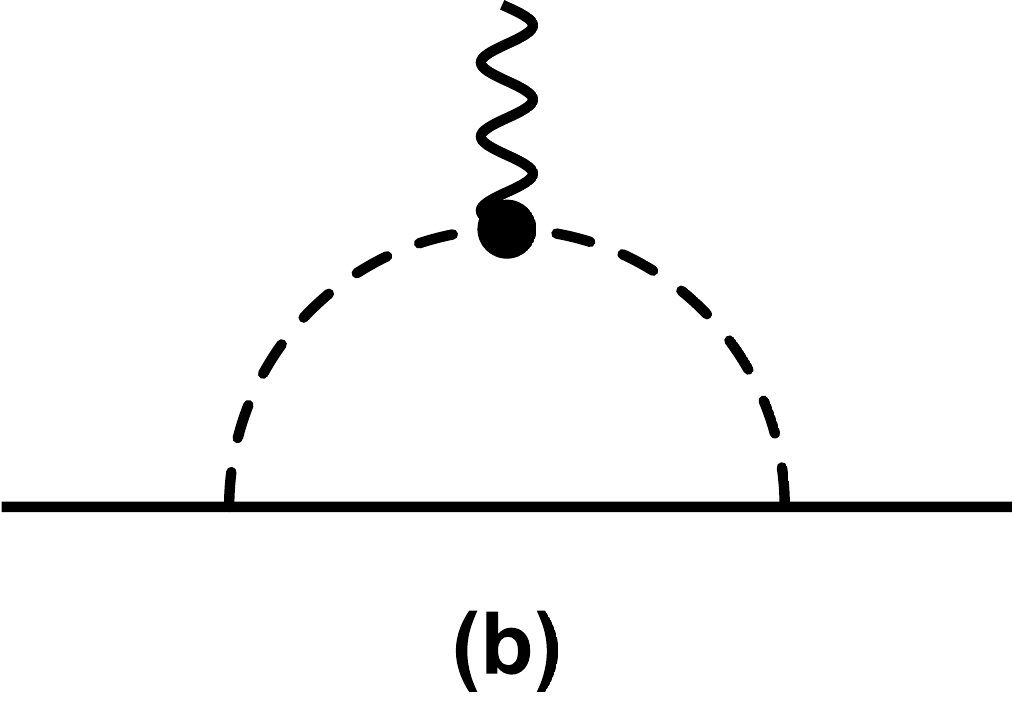}~~~\includegraphics[width=4cm]{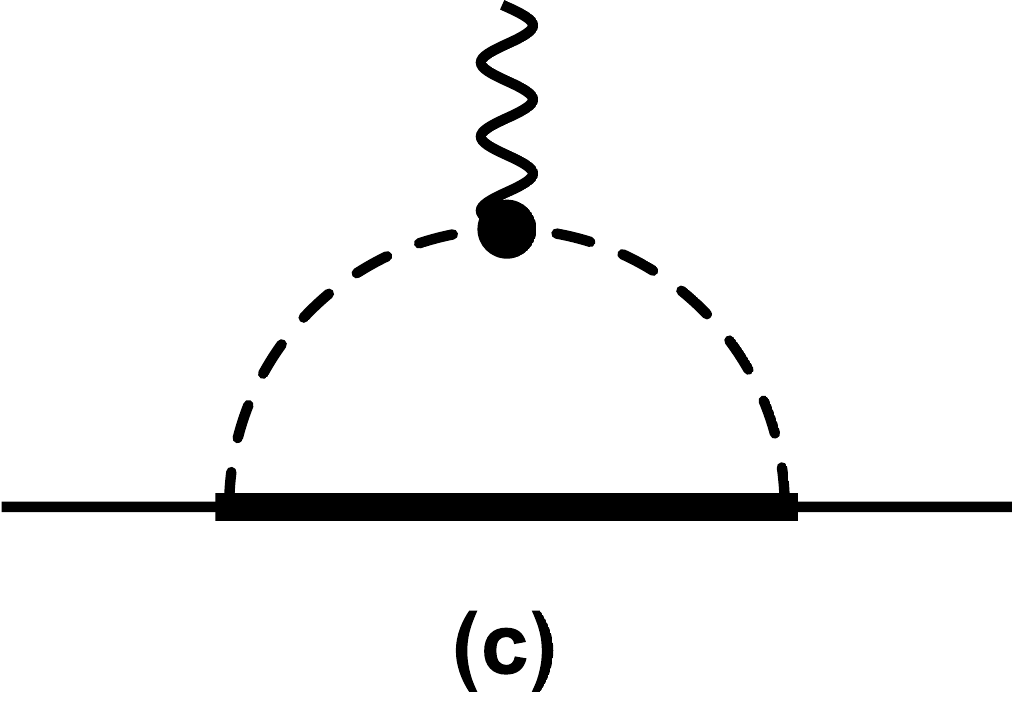}\\
  \includegraphics[width=4cm]{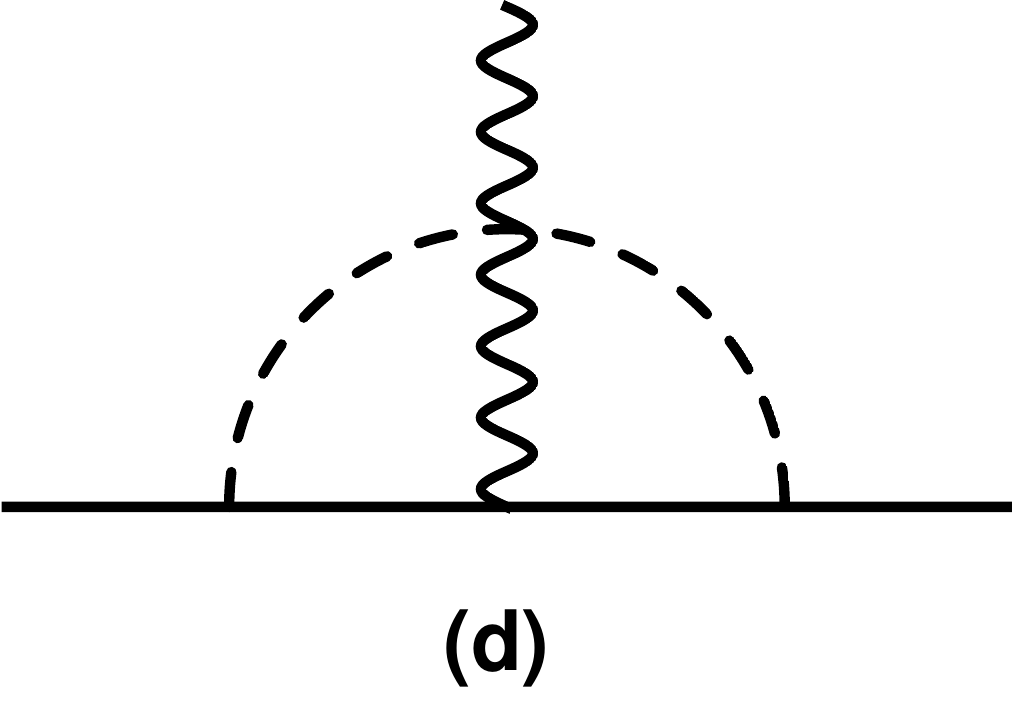}~~~\includegraphics[width=4cm]{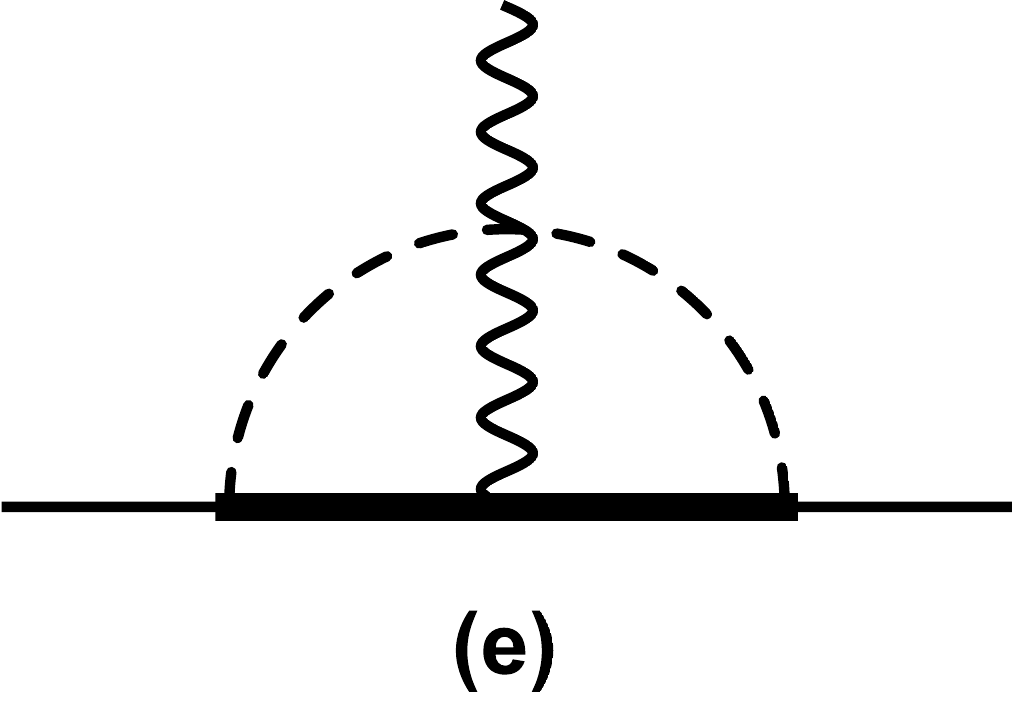}\\
  \caption{Feynman diagrams contributing to the singly charmed baryon magnetic moments up to NLO. Diagram (a) contributes at LO, while the other diagrams contribute at NLO. The solid, dashed, and wiggly lines represent singly charmed baryon, Goldstone bosons, and photons, respectively. The heavy dots denote the ${\cal O}(p^2)$ vertices.}\label{Fig1}
\end{figure}

The five Feynman diagrams contributing to $\mu_B$ up to ${\cal O}(p^3)$ are shown in Fig.1. The leading order contribution of ${\cal O}(p^2)$  is provided by the following Lagrangian:
\begin{eqnarray}
{\cal L}_{33}^{(2)}&=&\frac{d_2}{16m_{\bar{3}}}{\rm Tr}(\bar{B}_{\bar{3}}\sigma^{\mu\nu}F_{\mu\nu}^+B_{\bar{3}})\nonumber\\
&&+\frac{d_3}{16m_{\bar{3}}}{\rm Tr}(\bar{B}_{\bar{3}}\sigma^{\mu\nu}B_{\bar{3}}){\rm Tr}(F_{\mu\nu}^+),\nonumber\\
{\cal L}_{66}^{(2)}&=&\frac{d_5}{8m_6}{\rm Tr}(\bar{B}_6\sigma^{\mu\nu}F_{\mu\nu}^+B_6)\nonumber\\
&&+\frac{d_6}{8m_6}{\rm Tr}(\bar{B}_6\sigma^{\mu\nu}B_6){\rm Tr}(F_{\mu\nu}^+),\label{eq:treeL}
\end{eqnarray}
where the numbers in the superscript are the chiral order, $\sigma^{\mu\nu}=\frac{i}{2}[\gamma^\mu,\gamma^\nu]$, $F_{\mu\nu}^+=|e|(u^\dag Q_hF_{\mu\nu}u+uQ_hF_{\mu\nu}u^\dag)$, $F_{\mu\nu}=\partial_\mu A_\nu-\partial_\nu A_\mu$, and $Q_h={\rm diag}(1,0,0)$ is the charge operator of the charmed baryon, $u={\rm exp}[i\Phi/2F_\phi]$, with the unimodular matrix containing the pseudoscalar
nonet, and $F_\phi$ the pseudoscalar decay constant. In the following analysis, we take $F_\pi=92.4~{\rm MeV}$, $F_K=1.22F_\pi$, and $F_\eta=1.3F_\pi$. In the SU(3) flavor representation, there are  three kinds  of singly charmed baryons, which are denoted as $B_{\bar{3}}$, $B_6$, and $B_6^{*\mu}$, respectively,
\begin{eqnarray*}
B_{\bar{3}}=\left(
\begin{array}{ccc}
0 & \Lambda_c^+ & \Xi_c^+\\
-\Lambda_c^+ & 0 & \Xi_c^0\\
-\Xi_c^+ & -\Xi_c^0 & 0
\end{array}
\right),
\end{eqnarray*}
\begin{eqnarray}
B_6=\left(
\begin{array}{ccc}
\Sigma_c^{++} & \frac{\Sigma_c^+}{\sqrt{2}} & \frac{\Xi_c^{'+}}{\sqrt{2}}\\
\frac{\Sigma_c^+}{\sqrt{2}} & \Sigma_c^0 & \frac{\Xi_c^{'0}}{\sqrt{2}}\\
\frac{\Xi_c^{'+}}{\sqrt{2}} & \frac{\Xi_c^{'0}}{\sqrt{2}} & \Omega_c^0
\end{array}
\right),\qquad B_6^{*\mu}=\left(
\begin{array}{ccc}
\Sigma_c^{*++} & \frac{\Sigma_c^{*+}}{\sqrt{2}} & \frac{\Xi_c^{*+}}{\sqrt{2}}\\
\frac{\Sigma_c^{*+}}{\sqrt{2}} & \Sigma_c^{*0} & \frac{\Xi_c^{*0}}{\sqrt{2}}\\
\frac{\Xi_c^{*+}}{\sqrt{2}} & \frac{\Xi_c^{*0}}{\sqrt{2}} & \Omega_c^{*0}
\end{array}
\right).
\end{eqnarray}
The spin of  the $B_{\bar{3}}$ and $B_6$ states is $1/2$ while the spin of
the $B_6^{*\mu}$ states is $3/2$.

In the numerical analysis, we take the average of the masses for each flavor multiplet, i.e., $m_{\bar{3}}=2408$~MeV, $m_6=2535$~MeV, and $m_{6^*}=2602$~MeV~\cite{Tanabashi:2018oca}. The mass differences are $\delta_1=m_6-m_{\bar{3}}=127~{\rm MeV}$, $\delta_2=m_{6^*}-m_{\bar{3}}=194~{\rm MeV}$, and $\delta_3=m_{6^*}-m_6=67~{\rm MeV}$.

The loop diagrams arising at NLO are determined in terms of the lowest order LECs from ${\cal L}_B^{(1)}$+${\cal L}_{MB}^{(1)}$+${\cal L}_M^{(2)}$, which are,
\begin{eqnarray}
{\cal L}_B^{(1)}&=&\frac{1}{2}{\rm Tr}[\bar{B}_{\bar{3}}(i\slashed{D}-m_{\bar{3}})B_{\bar{3}}]+{\rm Tr}[\bar{B}_6(i\slashed{D}-m_6)B_6]\nonumber\\
&&+{\rm Tr}[\bar{B}_6^{*\mu}(-g_{\mu\nu}(i\slashed{D}-m_{6^*})+i(\gamma_\mu D_\nu+\gamma_\nu D_\mu)\nonumber\\
&&-\gamma_\mu(i\slashed{D}+m_{6^*})\gamma_\nu B_6^{*\nu}],\nonumber\\
{\cal L}_{MB}^{(1)}&=&\frac{g_1}{2}{\rm Tr}[\bar{B}_6\slashed{u}\gamma_5B_6]+\frac{g_2}{2}{\rm Tr}[\bar{B}_6\slashed{u}\gamma_5B_{\bar{3}}+{\rm h.c.}]\nonumber\\
&&+\frac{g_3}{2}{\rm Tr}[\bar{B}_6^{*\mu}u_\mu B_6+{\rm h.c.}]+\frac{g_4}{2}{\rm Tr}[\bar{B}_6^{*\mu}u_\mu B_{\bar{3}}+{\rm h.c.}]\nonumber\\
&&+\frac{g_5}{2}{\rm Tr}[\bar{B}_6^{*\mu}\slashed{u}\gamma_5B_{6\mu}^*]+\frac{g_6}{2}{\rm Tr}[\bar{B}_{\bar{3}}\slashed{u}\gamma_5B_{\bar{3}}],\nonumber\\
{\cal L}_M^{(2)}&=&\frac{F_\phi^2}{4}{\rm Tr}[\nabla_\mu U(\nabla^\mu U)^\dag],
\end{eqnarray}
with
\begin{eqnarray}
&&D_\mu B=\partial_\mu B+\Gamma_\mu B+B\Gamma_\mu^T,\nonumber\\
&&\Gamma_\mu=\frac{1}{2}(u^\dag\partial_\mu u+u\partial_\mu u^\dag)-\frac{i}{2}(u^\dag v_\mu u+uv_\mu u^\dag)=-ieQ_hA_\mu,\nonumber\\
&&u_\mu=i(u^\dag\partial_\mu u-u\partial_\mu u^\dag)+(u^\dag v_\mu u-uv_\nu u^\dag),\nonumber\\
&&U=u^2=e^{\frac{i\Phi}{F_\phi}},\qquad\nabla_\mu U=\partial_\mu U+ieA_\mu[Q_l,U],
\end{eqnarray}
where $v_\mu$ stands for the vector source, and the charge matrix for the light quark is $Q_l={\rm diag}(2/3,-1/3,-1/3)$.

The total spin of the light quarks is 0 for the singly charmed baryon in the $B_{\bar{3}}$ state. Considering parity and angular momentum conservation, the $B_{\bar{3}}B_{\bar{3}}\phi$ vertex is forbidden, i.e., $g_6=0$.

For the $B_{\bar{3}}$ and $B_6$ states, the tree level contributions of the magnetic moments can be easily obtained from Eq.~(\ref{eq:treeL}), which are:
\begin{eqnarray}
&&\kappa_{\bar{3}}^{(a,2)}=\alpha_{\bar{3}}d_2+\beta_{\bar{3}}d_3,\nonumber\\
&&\kappa_6^{(a,2)}=\alpha_6d_5+\beta_6d_6.\label{eq:TreeCCs3f}
\end{eqnarray}
The values of $\alpha_{\bar{3}}$, $\beta_{\bar{3}}$, $\alpha_6$, and $\beta_6$ are tabulated in  Table~\ref{tab:1} and Table~\ref{tab:2}. The four LECs $d_2$, $d_3$, $d_5$ and $d_6$ will be determined by fitting to lattice QCD data.
\begin{table}[htb]
 \caption{\label{tab:1}Coefficients of the tree level contributions of Eq.~(\ref{eq:TreeCCs3f}) for the $B_{\bar{3}}$ states.}
\begin{center}
    \begin{tabular}{cccc}
      \hline
      \hline
      ~~~~~~ & ~~~~~~$\Lambda_c^+$~~~~~~ & ~~~~~~$\Xi_c^+$~~~~~~ & ~~~~~~$\Xi_c^0$~~~~~~\\
      \hline
      $\alpha_{\bar{3}}$ & $\frac{1}{2}$ & $\frac{1}{2}$ & $0$\\
      \hline
      $\beta_{\bar{3}}$ & $1$ & $1$ & $1$\\
      \hline
      \hline
    \end{tabular}
  \end{center}
\end{table}

At ${\cal O}(p^3)$, the loop contributions to the magnetic moments, which come from diagrams (b), (c), (d), and (e) in Fig.~\ref{Fig1}, are written as,
\begin{eqnarray}
\kappa_{\bar{3}}^{(3)}&=&\frac{1}{4\pi^2}
\left(\sum_{\phi=\pi,K}\frac{g_2^2}{F_\phi^2}\xi_{B_{\bar{3}}\phi,\delta_1}^{(3,b)}H_{B_{\bar{3}}}^{(b)}(\delta_1,m_\phi)\right.\nonumber\\
&&+\sum_{\phi=\pi,K}\frac{g_4^2}{F_\phi^2}\xi_{B_{\bar{3}}\phi,\delta_2}^{(3,c)}H_{B_{\bar{3}}}^{(c)}(\delta_2,m_\phi)\nonumber\\
&&+\sum_{\phi=\pi,K,\eta}\frac{g_2^2}{F_\phi^2}\xi_{B_{\bar{3}}\phi,\delta_1}^{(3,d)}H_{B_{\bar{3}}}^{(d)}(\delta_1,m_\phi)\nonumber\\
&&\left.+\sum_{\phi=\pi,K,\eta}\frac{g_4^2}{F_\phi^2}\xi_{B_{\bar{3}}\phi,\delta_2}^{(3,e)}H_{B_{\bar{3}}}^{(e)}(\delta_2,m_\phi)\right),\nonumber\\
\kappa_6^{(3)}&=&\frac{1}{4\pi^2}\left(\sum_{\phi=\pi,K}\frac{g_1^2}{F_\phi^2}\xi_{B_6\phi}^{(3,b)}H_{B_6}^{(b)}(0,m_\phi)\right.\nonumber\\
&&+\sum_{\phi=\pi,K}\frac{g_2^2}{F_\phi^2}\xi_{B_6\phi,\delta_1}^{(3,b)}H_{B_6}^{(b)}(\delta_1,m_\phi)\nonumber\\
&&+\sum_{\phi=\pi,K}\frac{g_3^2}{F_\phi^2}\xi_{B_6\phi,\delta_3}^{(3,c)}H_{B_6}^{(c)}(\delta_3,m_\phi)\nonumber\\
&&+\sum_{\phi=\pi,K,\eta}\frac{g_1^2}{F_\phi^2}\xi_{B_6\phi}^{(3,d)}H_{B_6}^{(d)}(0,m_\phi)\nonumber\\
&&+\sum_{\phi=\pi,K,\eta}\frac{g_2^2}{F_\phi^2}\xi_{B_6\phi,\delta_1}^{(3,d)}H_{B_6}^{(d)}(\delta_1,m_\phi)\nonumber\\
&&\left.+\sum_{\phi=\pi,K,\eta}\frac{g_3^2}{F_\phi^2}\xi_{B_6\phi,\delta_3}^{(3,e)}H_{B_6}^{(e)}(\delta_3,m_\phi)\right),\label{eq:LoopCCs3f}
\end{eqnarray}
with the coefficients $\xi_{B_{\bar{3}}\phi,\delta_i}^{(3;b,c,d,e)}$, $\xi_{B_6\phi,\delta_i}^{(3;b,c,d,e)}$ listed in Table~\ref{tab:3} and Table~\ref{tab:4}. The explicit expressions of the loop functions $H_{B_{\bar{3}}}^{(b,c,d,e)}(\delta_i,m_\phi)$ and $H_{B_6}^{(b,c,d,e)}(\delta_i,m_\phi)$ can be found in the Appendix.

Once we obtain the loop functions in the EOMS scheme, we can easily obtain their HB counterparts by performing $1/m_0$ expansions, We have
checked that our results agree with those of Ref.~\cite{Wang:2018xoc}. In the following section, for the sake of comparison, we study also the performance of
the HBChPT in describing the lattice QCD data of Refs.~\cite{Bahtiyar:2016dom,Can:2013tna,Can:2015exa}.  It should be noted that in the following section, unless otherwise stated, the HBChPT results
refer to the ones obtained in the present work, not those of Ref.~\cite{Wang:2018xoc}

\begin{table}[htb]
 \caption{\label{tab:2}Coefficients of the tree level contributions of Eq.~(\ref{eq:TreeCCs3f}) for the $B_6$ states.}
\begin{center}
    \begin{tabular}{ccccccc}
      \hline
      \hline
      ~~~~~~ & ~~~$\Sigma_c^{++}$~~~ & ~~~$\Sigma_c^+$~~~ & ~~~$\Sigma_c^0$~~~ & ~~~$\Xi_c^{'+}$~~~ & ~~~$\Xi_c^{'0}$~~~ & ~~~$\Omega_c^0$~~~\\
      \hline
      $\alpha_6$ & $1$ & $\frac{1}{2}$ & $0$ & $\frac{1}{2}$ & $0$ & $0$\\
      \hline
      $\beta_6$ & $1$ & $1$ & $1$ & $1$ & $1$ & $1$\\
      \hline
      \hline
    \end{tabular}
  \end{center}
\end{table}
\begin{table}[h!]
 \caption{\label{tab:3}Coefficients of the loop contributions of Eq.~(\ref{eq:LoopCCs3f}) for the $B_{\bar{3}}$ states.}
\begin{center}
    \begin{tabular}{cccc}
      \hline
      \hline
      ~~~~~~ & ~~~~~~$\Lambda_c^+$~~~~~~ & ~~~~~~$\Xi_c^+$~~~~~~ & ~~~~~~$\Xi_c^0$~~~~~~\\
      \hline
      $\xi_{B_{\bar{3}}\pi,\delta_1}^{(3,b)}$ & $0$ & $1$ & $-1$\\
      \hline
      $\xi_{B_{\bar{3}}K,\delta_1}^{(3,b)}$ & $1$ & $0$ & $-1$\\
      \hline
      $\xi_{B_{\bar{3}}\pi,\delta_2}^{(3,c)}$ & $0$ & $1$ & $-1$\\
      \hline
      $\xi_{B_{\bar{3}}K,\delta_2}^{(3,c)}$ & $1$ & $0$ & $-1$\\
      \hline
      $\xi_{B_{\bar{3}}\pi,\delta_1}^{(3,d)}$ & $6$ & $\frac{1}{2}$ & $1$\\
      \hline
      $\xi_{B_{\bar{3}}K,\delta_1}^{(3,d)}$ & $1$ & $5$ & $1$\\
      \hline
      $\xi_{B_{\bar{3}}\eta,\delta_1}^{(3,d)}$ & $0$ & $\frac{3}{2}$ & $0$\\
       \hline
      $\xi_{B_{\bar{3}}\pi,\delta_2}^{(3,e)}$ & $3$ & $\frac{1}{4}$ & $\frac{1}{2}$\\
      \hline
      $\xi_{B_{\bar{3}}K,\delta_2}^{(3,e)}$ & $\frac{1}{2}$ & $\frac{5}{2}$ & $\frac{1}{2}$\\
      \hline
      $\xi_{B_{\bar{3}}\eta,\delta_2}^{(3,e)}$ & $0$ & $\frac{3}{4}$ & $0$\\
      \hline
      \hline
    \end{tabular}
  \end{center}
\end{table}
\begin{table}[h!]
 \caption{\label{tab:4}Coefficients of the loop contributions of Eq.~(\ref{eq:LoopCCs3f}) for the $B_6$ states.}
\begin{center}
    \begin{tabular}{ccccccc}
      \hline
      \hline
       ~~~~~~ & ~~~$\Sigma_c^{++}$~~~ & ~~~$\Sigma_c^+$~~~ & ~~~$\Sigma_c^0$~~~ & ~~~$\Xi_c^{'+}$~~~ & ~~~$\Xi_c^{'0}$~~~ & ~~~$\Omega_c^0$~~~\\
      \hline
      $\xi_{B_6\pi}^{(3,b)}$ & $1$ & $0$ & $-1$ & $\frac{1}{2}$ & $-\frac{1}{2}$ & $0$\\
      \hline
      $\xi_{B_6K}^{(3,b)}$ & $1$ & $\frac{1}{2}$ & $0$ & $0$ & $-\frac{1}{2}$ & $$-1\\
      \hline
      $\xi_{B_6\pi,\delta_1}^{(3,b)}$ & $2$ & $0$ & $-2$ & $1$ & $-1$ & $0$\\
      \hline
      $\xi_{B_6K,\delta_1}^{(3,b)}$ & $2$ & $1$ & $0$ & $0$ & $-1$ & $-2$\\
      \hline
      $\xi_{B_6\pi,\delta_3}^{(3,c)}$ & $1$ & $0$ & $-1$ & $\frac{1}{2}$ & $-\frac{1}{2}$ & $0$\\
      \hline
      $\xi_{B_6K,\delta_3}^{(3,c)}$ & $1$ & $\frac{1}{2}$ & $0$ & $0$ & $-\frac{1}{2}$ & $$-1\\
       \hline
      $\xi_{B_6\pi}^{(3,d)}$ & $3$ & $2$ & $1$ & $\frac{1}{4}$ & $\frac{1}{2}$ & $0$\\
      \hline
      $\xi_{B_6K}^{(3,d)}$ & $1$ & $\frac{1}{2}$ & $0$ & $\frac{5}{2}$ & $\frac{1}{2}$ & $1$\\
      \hline
      $\xi_{B_6\eta}^{(3,d)}$ & $\frac{2}{3}$ & $\frac{1}{3}$ & $0$ & $\frac{1}{12}$ & $0$ & $0$\\
      \hline
      $\xi_{B_6\pi,\delta_1}^{(3,d)}$ & $2$ & $2$ & $2$ & $\frac{1}{2}$ & $1$ & $0$\\
      \hline
      $\xi_{B_6K,\delta_1}^{(3,d)}$ & $2$ & $1$ & $0$ & $1$ & $1$ & $2$\\
      \hline
      $\xi_{B_6\eta,\delta_1}^{(3,d)}$ & $0$ & $0$ & $0$ & $\frac{3}{2}$ & $0$ & $0$\\
      \hline
      $\xi_{B_6\pi,\delta_3}^{(3,e)}$ & $\frac{3}{2}$ & $1$ & $\frac{1}{2}$ & $\frac{1}{8}$ & $\frac{1}{4}$ & $0$\\
      \hline
      $\xi_{B_6K,\delta_3}^{(3,e)}$ & $\frac{1}{2}$ & $\frac{1}{4}$ & $0$ & $\frac{5}{4}$ & $\frac{1}{4}$ & $\frac{1}{2}$\\
      \hline
      $\xi_{B_6\eta,\delta_3}^{(3,e)}$ & $\frac{1}{3}$ & $\frac{1}{6}$ & $0$ & $\frac{1}{24}$ & $0$ & $0$\\
      \hline
      \hline
    \end{tabular}
  \end{center}
\end{table}

\section{Results and discussions}
\begin{table*}[t]
 \caption{\label{tab:5}Magnetic moments of singly charmed baryons at different $m_\pi$~\cite{Bahtiyar:2016dom,Can:2013tna,Can:2015exa,Bahtiyar:2015sga}, in units of nuclear magneton~[$\mu_N$].}
\begin{center}
    \begin{tabular}{cccccccc}
      \hline
      \hline
      ~~~$m_\pi$~(MeV)~~~ & ~~~$\Xi_c^+$~~~  & ~~~$\Xi_c^0$~~~  & ~~~$\Sigma_c^{++}$~~~ & ~~~$\Sigma_c^0$~~~ & ~~~$\Xi_c^{'+}$~~~ & ~~~$\Xi_c^{'0}$~~~ & ~~~$\Omega_c^0$~~~\\
      \hline
      ~~~Phys.~~~ & $\cdots$ & $\cdots$ & $1.499(202)$ & $-0.875(103)$ & $\cdots$ & $\cdots$ & $-0.667(96)$\\
      \hline
      ~~~$156$~~~ & $0.235(25)$ & $0.192(17)$ & $\cdots$ & $\cdots$ & $0.315(141)$ & $-0.599(71)$ & $-0.688(31)$\\
      \hline
      ~~~$300$~~~ & $\cdots$ & $\cdots$ & $1.867(388)$ & $-0.929(206)$ & $\cdots$ & $\cdots$ & $-0.640(55)$\\
      \hline
      ~~~$410$~~~ & $\cdots$ & $\cdots$ & $1.591(358)$ & $-0.897(223)$ & $\cdots$ & $\cdots$ & $-0.621(44)$\\
      \hline
      ~~~$570$~~~ & $\cdots$ & $\cdots$ & $1.289(161)$ & $-0.724(80)$ & $\cdots$ & $\cdots$ & $-0.658(46)$\\
      \hline
      ~~~$700$~~~ & $\cdots$ & $\cdots$ & $1.447(125)$ & $-0.757(67)$ & $\cdots$ & $\cdots$ & $-0.701(56)$\\
      \hline
      \hline
    \end{tabular}
  \end{center}
\end{table*}

In this section, we determine the LECs $d_2$, $d_3$, $d_5$ and $d_6$ by fitting to the lattice QCD data of Refs.~\cite{Bahtiyar:2016dom,Can:2013tna,Can:2015exa}, which are collected in Table~\ref{tab:5} for the sake of easy reference. Because of the limited lattice QCD data, the other LECs $g_{1-4}$ are fixed by the quark model and the heavy quark spin flavor symmetry. Their values are $g_1=0.98$, $g_2=-\sqrt{\frac{3}{8}}g_1=-0.60$, $g_3=\frac{\sqrt{3}}{2}g_1=0.85$, and $g_4=-\sqrt{3}g_2=1.04$~\cite{Jiang:2015xqa,Jiang:2015xqa,Jiang:2014ena}. In our least-squares fit, the $\chi^2$ as a function of the LECs is defined as
\begin{eqnarray}
\chi^2({\rm C_X})=\sum_{i=1}^n\frac{(\mu_i^{\rm th}({\rm C_X})-\mu_i^{\rm LQCD})^2}{\sigma_i^2},
\end{eqnarray}
where $C_X$ denote all the LECs, $\sigma_i$ correspond to the uncertainty of each lattice QCD datum, $\mu_i^{\rm th}({\rm C_X})$ and $\mu_i^{\rm LQCD}$ stand for the magnetic moments obtained in the BChPT and those of the lattice QCD in Table~\ref{tab:5}, respectively.

In order to decompose the contributions of loop diagrams, we will consider two cases. In case 1, all the allowed intermediate baryons are taken into account, while in case 2, only intermediate baryons of the same type as those of the external baryons are
considered.   Fitting to the lattice QCD data of Table ~\ref{tab:5} and with $g_{1-4}$ fixed, the resulting LECs and $\chi^2$ are listed in Table~\ref{tab:6}. One notes
that the EOMS BChPT descriptions of the lattice QCD data are better than that of the HB BChPT in both cases.
\begin{table}[h!]
 \caption{\label{tab:6} LECs $d_2$, $d_3$, $d_5$, and $d_6$ determined by fitting to the lattice QCD data, with $g_{1-4}$ fixed.
 In case 1 all the allowed intermediate baryons in the loop diagrams are taken into account, while in case 2
 only intermediate baryons of the same type as those of  the external baryons in the loop diagrams are considered.  }
\begin{center}
    \begin{tabular}{cccccc}
      \hline
      \hline
      \multirow{2}{0.5cm}  &  \multicolumn{2}{c}{Case 1} &  & \multicolumn{2}{c}{Case 2}\\
      \cline{2-3}\cline{5-6}
      &  EOMS 1  & HB 1 & & EOMS 2 & HB 2\\
      \hline
      $d_2$ & $-1.25(15)$ & $-2.32(15)$ & & $-1.78(15)$ & $-1.78(15)$ \\
      \hline
      $d_3$ & $2.20(4)$ & $0.65(4)$ & & $0.49(4)$ & $0.49(4)$ \\
      \hline
      $d_5$ & $7.83(34)$ & $13.49(34)$ & & $5.08(34)$ & $8.69(34)$ \\
      \hline
      $d_6$ & $-3.76(5)$ & $-4.93(5)$ & & $-2.66(5)$ & $-3.40(5)$ \\
      \hline
      $g_1$ & $0.98$ & $0.98$ & & $0.98$ & $0.98$ \\
      \hline
      $g_2$ & $-0.60$ & $0$ & & $-0.60$ & $0$ \\
      \hline
      $g_3$ & $0.85$ & $0.85$ & & $0.85$ & $0.85$ \\
      \hline
      $g_4$ & $1.04$ & $0$ & & $1.04$ & $0$ \\
      \hline
      $\chi_{\rm min}^2$ & $41.42$ & $131.05$ & & $15.10$ & $34.35$ \\
      \hline
      \hline
    \end{tabular}
  \end{center}
\end{table}

 Note that we do not fit to the lattice QCD data obtained at $m_\pi=700$~MeV, which are probably out of the range of validity of NLO ChPT. Furthermore, as can be seen in Fig.~\ref{Fig3},
the difference between the lattice QCD value and the ChPT prediction for $\mu_{\Xi_c^{'0}}$ is somehow relatively large. Thus, we do not include the lattice QCD magnetic moment
 of $\Xi_c^{'0}$ in our fitting as well.

\begin{figure*}[h!]
  \centering
  \includegraphics[width=16cm]{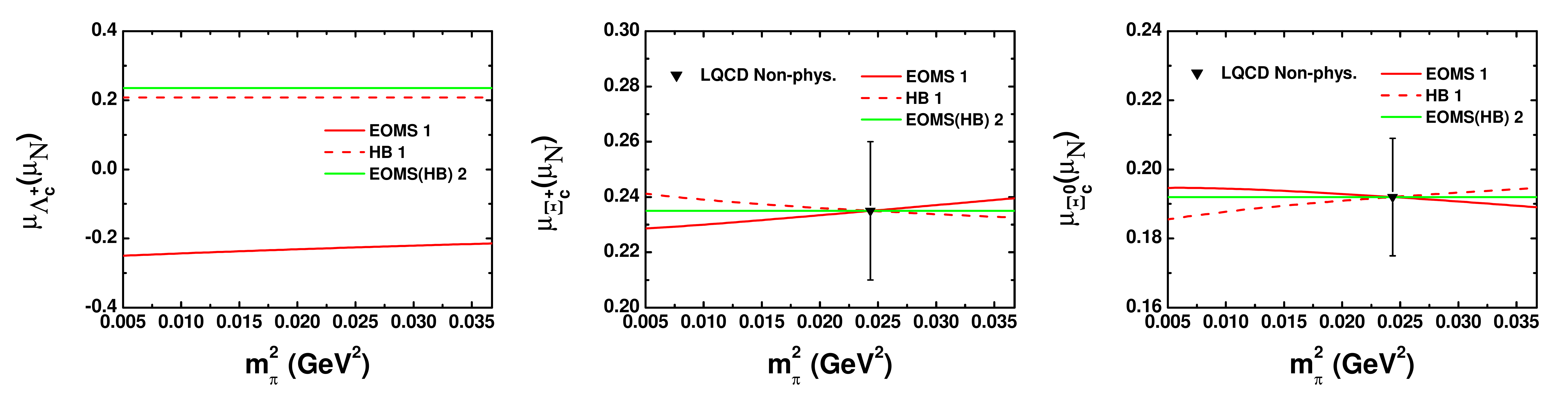}\\
  \caption{Magnetic moments of the singly charmed anti-triplet baryons  as a function of $m_\pi^2$. The solid black nablas represent the corresponding lattice QCD data that are fitted.}\label{Fig2}
\end{figure*}

\begin{figure*}[h!]
  \centering
  \includegraphics[width=16cm]{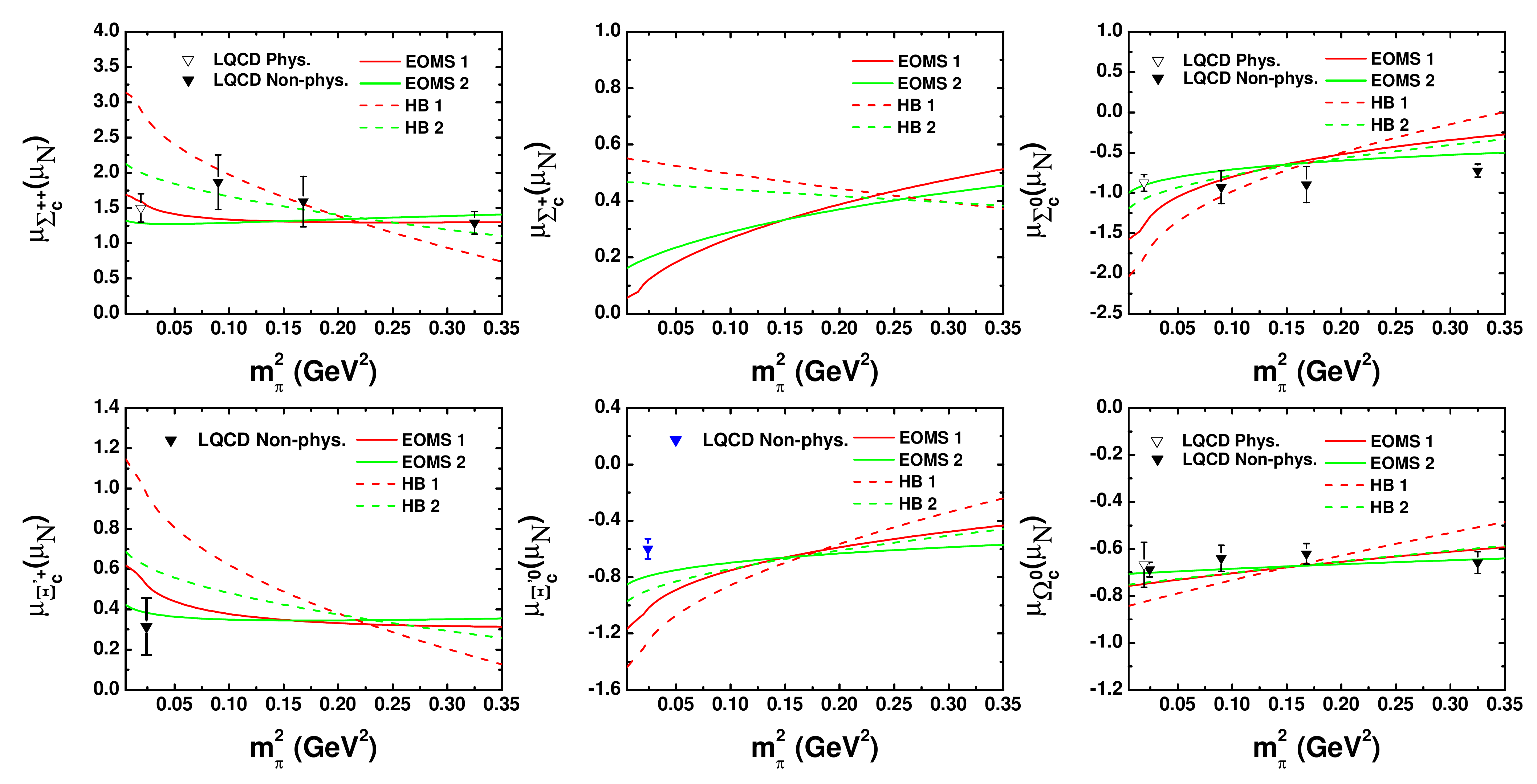}\\
  \caption{Magnetic moments of the singly charmed sextet baryons as a function of $m_\pi^2$. The solid black nablas refer to the corresponding  lattice QCD data  fitted. The hollow nablas stand for the lattice QCD physical values. The blue nablas denote the lattice QCD data not used in our fitting.}\label{Fig3}
\end{figure*}

 For the sake of comparison with the lattice QCD data, in Fig.~\ref{Fig2}, we present the predicted magnetic moments of the singly charmed anti-triplet baryons as a function of $m_\pi^2$. It is seen that the EOMS BChPT results are of the same qualify as those of the HB BChPT for  $\Xi_c^+$ and $\Xi_c^0$. However, surprisingly,  the EOMS and HB  predictions for $\Lambda_c^+$ in case 1 are very different. From Table~\ref{tab:7}, we note that in the HBChPT the contributions from the intermediate anti-triplet and sextet baryons cancel each other at ${\cal O}(p^3)$. Thus, at this order, loop corrections are quite small. But in the EOMS scheme, the loop contributions are rather large, especially for $\Lambda_c^+$. In addition, we note that the main contributions of the loop diagrams are from the baryon pole diagram. Therefore, the large difference for the prediction of $\mu_{\Lambda_c^+}$ is caused
 by the absence of  the baryon pole diagram in the HB BChPT at $\mathcal{O}(p^3)$.

 In Fig.~\ref{Fig3}, we plot the predicted magnetic moments of the singly charmed sextet  baryons  as a function of $m_\pi^2$, in comparison with the lattice QCD data.  The EOMS BChPT  results  are in better agreement with the
 lattice QCD data than those of the HB BChPT. As shown in Tables \ref{tab:6},  on average the description of the lattice QCD data becomes worse if the intermediate anti-triplet states are
 included. Therefore, on average the results obtained in case 2 are  in better agreement with the lattice QCD data. This has been noted in Ref.~\cite{Wang:2018xoc} as well.
In Tables \ref{tab:7} and \ref{tab:8},  we decompose the loop contributions mediated by the $\bar{3}$, 6, and $6^*$ states. One can see that the convergence pattern in case 2 is in generally better than that in case 1, with probably the exception of $\Sigma^0_c$. Therefore, we take the predictions obtained in case 2 as our final results.

\begin{figure*}[h!]
  \centering
  \includegraphics[width=12cm]{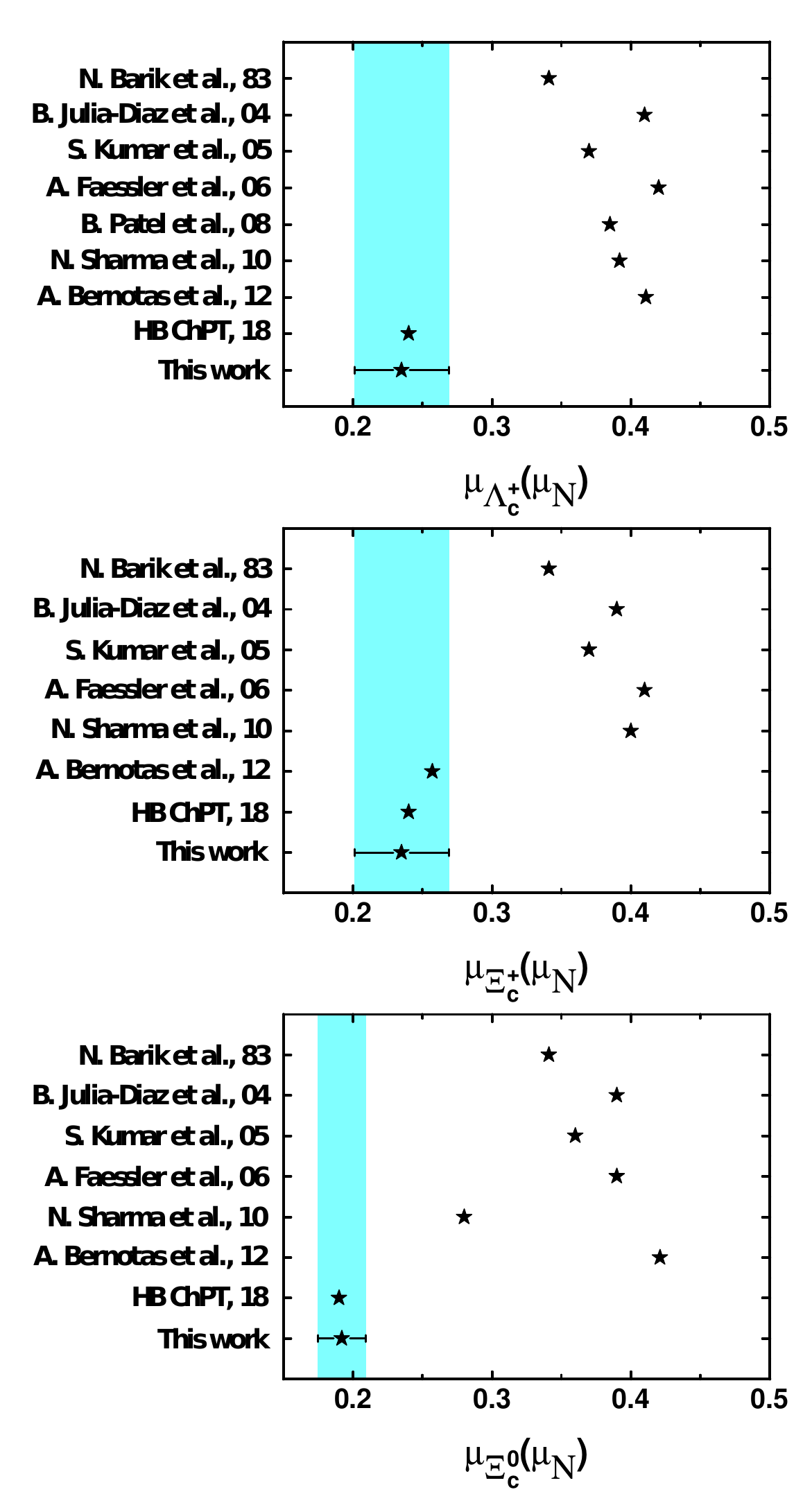}\\
  \caption{Magnetic moments of the anti-triplet baryons obtained in different approaches. The light-blue bands represent the result obtained in the present work. The others are taken from Ref.~\cite{Barik:1984tq} (N. Barik et al., 83), Ref.~\cite{JuliaDiaz:2004vh} (B. Julia-Diaz et al., 04), Ref.~\cite{Kumar:2005ei} (S. Kumar et al., 05), Ref.~\cite{Faessler:2006ft} (A. Faessler et al., 06), Ref.~\cite{Patel:2007gx} (B. Patel et al., 08), Ref.~\cite{Sharma:2010vv} (N. Sharma et al., 10), Ref.~\cite{Bernotas:2012nz} (A. Bernotas et al., 12), and Ref.~\cite{Wang:2018xoc} (HB ChPT, 18).}\label{Fig4}
\end{figure*}

\begin{figure*}[h!]
  \centering
  \includegraphics[width=16cm]{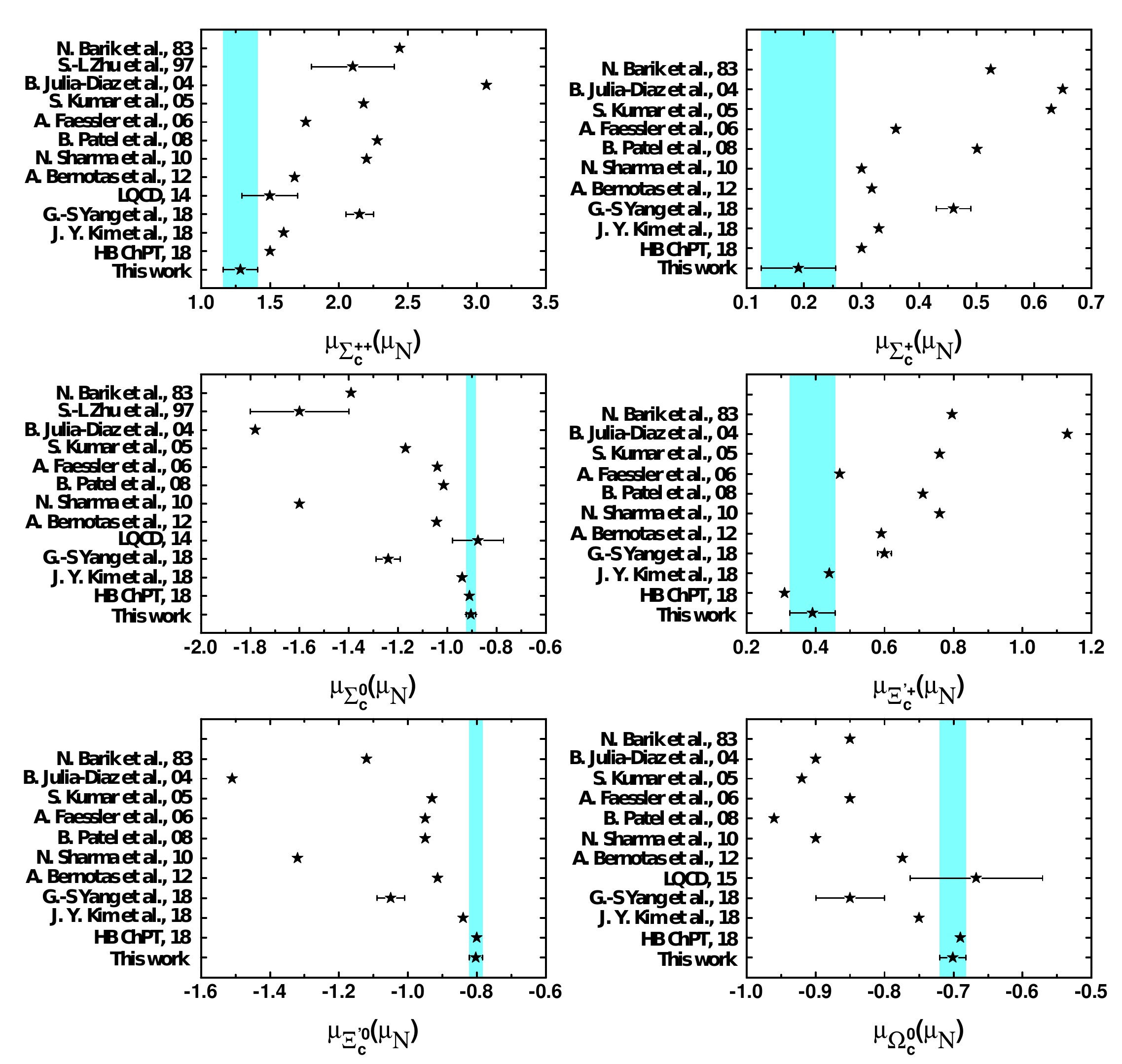}\\
  \caption{ Same as Fig. 4, but for the sexet baryons. Additional data are taken from Ref.~\cite{Zhu:1997as} (S.-L Zhu et al., 97), Ref.~\cite{Yang:2018uoj} (G.-S Yang et al., 18), Ref.~\cite{Kim:2018nqf} (J. Y. Kim et al., 18), Ref.~\cite{Can:2013tna} (LQCD, 14), and Ref.~\cite{Bahtiyar:2015sga} (LQCD, 15).}\label{Fig5}
  \vspace{5cm}
\end{figure*}

 In Fig.~\ref{Fig4} and Fig.~\ref{Fig5}, we compare the predicted magnetic moments of all the singly charmed baryons at the physical point with those obtained in other approaches. We note that the results of different approaches are rather scattered. However, our results are in better agreement with those of the HBChPT of Ref.~\cite{Wang:2018xoc}, though
 we have chosen different strategies to determine some of the LECs. Clearly,
 further experimental or lattice QCD studies are needed to pin down their values and to discriminate between different theoretical approaches.

\begin{table*}[htb]
 \caption{\label{tab:7}Decomposition of the loop contributions to the magnetic moments of singly charmed baryons. The subscript $\bar{3}$, $6$, and $6^*$ denote the loop diagrams with the intermediate $\bar{3}$,  6, and $6^*$ states at ${\cal O}(p^3)$, respectively.}
\begin{center}
    \begin{tabular}{ccccccccccccccc}
      \hline
      \hline
      \multirow{2}{0.5cm}  &  & \multicolumn{5}{c}{EOMS 1} &  & \multicolumn{5}{c}{HB 1} & & \multirow{2}{2cm}{\rm LQCD~\cite{Bahtiyar:2015sga,Can:2013tna}}\\
      \cline{3-7}\cline{9-13}
      &  & ${\cal O}(p^2)$ & ${\cal O}(p^3)_{\bar{3}}$ & ${\cal O}(p^3)_6$ & ${\cal O}(p^3)_{6^*}$ & $\mu_{\rm tot}$ & & ${\cal O}(p^2)$ & ${\cal O}(p^3)_{\bar{3}}$ & ${\cal O}(p^3)_6$ & ${\cal O}(p^3)_{6^*}$ & $\mu_{\rm tot}$ & &\\
      \hline
     \multirow{3}{0.5cm}{$B_{\bar{3}}$} & $\mu_{\Lambda_c^+}$ & $1.005$ & $\cdots$ & $0.035$ & $-1.272$ & $-0.232$ & $~~~$ & $0.191$ & $\cdots$ &$-0.263$ & $0.280$ & $0.208$ & & $\cdots$\\
      \cline{2-15}
      & $\mu_{\Xi_c^+}$ & $1.005$ & $\cdots$ & $0.141$ & $-0.913$ & $0.233$ &  & $0.191$ & $\cdots$ & $-0.169$ & $0.215$ & $0.237$ & & $\cdots$\\
      \cline{2-15}
      & $\mu_{\Xi_c^0}$ & $0.859$ & $\cdots$ & $0.330$ & $-0.996$ & $0.193$ &  & $0.253$ & $\cdots$ & $0.432$ & $-0.495$ & $0.190$ &  & $\cdots$\\
      \hline
      \multirow{6}{0.5cm}{$B_6$} & $\mu_{\Sigma_c^{++}}$ & $2.251$ & $-0.293$ & $-0.444$ & $0.090$ & $1.604$ &  & $3.916$ & $-0.319$ & $-0.988$ & $0.288$ & $2.897$ & & $1.499(202)$\\
      \cline{2-15}
      & $\mu_{\Sigma_c^+}$ & $0.428$ & $-0.192$ & $-0.094$ & $-0.042$ & $0.100$ &  & $1.044$ & $-0.243$ & $-0.349$ & $0.091$ & $0.543$ & & $\cdots$\\
      \cline{2-15}
      & $\mu_{\Sigma_c^0}$ & $-1.394$ & $-0.090$ & $0.256$ & $-0.175$ & $-1.403$ &  & $-1.828$ & $-0.168$ & $0.290$ & $-0.106$ & $-1.812$ & & $-0.875(103)$\\
      \cline{2-15}
      & $\mu_{\Xi_c^{'+}}$ & $0.428$ & $0.067$ & $0.112$ & $-0.048$ & $0.559$ &  & $1.044$ & $0.084$ & $-0.145$ & $0.053$ & $1.036$ & & $\cdots$\\
      \cline{2-15}
      & $\mu_{\Xi_c^{'0}}$ & $-1.394$ & $0.135$ & $0.380$ & $-0.198$ & $-1.077$ & & $-1.828$ & $0.159$ & $0.494$ & $-0.144$ & $-1.319$ & & $\cdots$\\
      \cline{2-15}
      & $\mu_{\Omega_c^0}$ & $-1.394$ & $0.361$ & $0.505$ & $-0.220$ & $-0.748$ &  & $-1.828$ & $0.486$ & $0.698$ & $-0.182$ & $-0.826$ & & $-0.667(96)$\\
      \hline
      \hline
    \end{tabular}
  \end{center}
\end{table*}
\begin{table*}[htb]
 \caption{\label{tab:8}Same as Table~\ref{tab:7} , but for  case 2.}
\begin{center}
    \begin{tabular}{ccccccccccccccc}
      \hline
      \hline
      \multirow{2}{0.5cm}  &  & \multicolumn{5}{c}{EOMS 2} &  & \multicolumn{5}{c}{HB 2} & & \multirow{2}{2cm}{\rm LQCD~\cite{Bahtiyar:2015sga,Can:2013tna}}\\
      \cline{3-7}\cline{9-13}
      &  & ${\cal O}(p^2)$ & ${\cal O}(p^3)_{\bar{3}}$ & ${\cal O}(p^3)_6$ & ${\cal O}(p^3)_{6^*}$ & $\mu_{\rm tot}$ & & ${\cal O}(p^2)$ & ${\cal O}(p^3)_{\bar{3}}$ & ${\cal O}(p^3)_6$ & ${\cal O}(p^3)_{6^*}$ & $\mu_{\rm tot}$ & &\\
      \hline
      \multirow{3}{0.5cm}{$B_{\bar{3}}$} & $\mu_{\Lambda_c^+}$ & $0.235$ & $\cdots$ & $\cdots$ & $\cdots$ & $0.235$ & $~~~$ & $0.235$ & $\cdots$ & $\cdots$ & $\cdots$ & $0.235$ & & $\cdots$\\
      \cline{2-15}
      & $\mu_{\Xi_c^+}$ & $0.235$ & $\cdots$ & $\cdots$ & $\cdots$ & $0.235$ &  & $0.235$ & $\cdots$ & $\cdots$ & $\cdots$ & $0.235$ & & $\cdots$\\
      \cline{2-15}
      & $\mu_{\Xi_c^0}$ & $0.192$ & $\cdots$ & $\cdots$ & $\cdots$ & $0.192$ &  & $0.192$ & $\cdots$ & $\cdots$ & $\cdots$ & $0.192$ & & $\cdots$\\
      \hline
      \multirow{6}{0.5cm}{$B_6$} & $\mu_{\Sigma_c^{++}}$ & $1.639$ & $\cdots$ & $-0.444$ & $0.090$ & $1.285$ &  & $2.703$ & $\cdots$ & $-0.988$ & $0.288$ & $2.003$ & & $1.499(202)$\\
      \cline{2-15}
      & $\mu_{\Sigma_c^+}$ & $0.326$ & $\cdots$ & $-0.094$ & $-0.042$ & $0.190$ &  & $0.721$ & $\cdots$ & $-0.349$ & $0.091$ & $0.463$ & & $\cdots$\\
      \cline{2-15}
      & $\mu_{\Sigma_c^0}$ & $-0.986$ & $\cdots$ & $0.256$ & $-0.175$ & $-0.905$ &  & $-1.261$ & $\cdots$ & $0.290$ & $-0.106$ & $-1.077$ & & $-0.875(103)$\\
      \cline{2-15}
      & $\mu_{\Xi_c^{'+}}$ & $0.326$ & $\cdots$ & $0.112$ & $-0.048$ & $0.390$ &  & $0.721$ & $\cdots$ & $-0.145$ & $0.053$ & $0.629$ & & $\cdots$\\
      \cline{2-15}
      & $\mu_{\Xi_c^{'0}}$ & $-0.986$ & $\cdots$ & $0.380$ & $-0.197$ & $-0.803$ &  & $-1.261$ & $\cdots$ & $0.494$ & $-0.144$ & $-0.911$ & & $\cdots$\\
      \cline{2-15}
      & $\mu_{\Omega_c^0}$ & $-0.986$ & $\cdots$ & $0.504$ & $-0.220$ & $-0.702$ &  & $-1.261$ & $\cdots$ & $0.698$ & $-0.182$ & $-0.745$ & & $-0.667(96)$\\
      \hline
      \hline
    \end{tabular}
  \end{center}
\end{table*}
\vspace{3cm}
\section{Summary}
Motivated by the recent experimental progress on heavy flavor hadrons, we have studied the magnetic moments of the singly charmed baryons in
the covariant baryon chiral perturbation theory (BChPT) up to the next-to-leading order. Using the quark model and the heavy quark spin flavor symmetry to fix some of the low energy constants, we determined the rest by fitting to the lattice QCD data.  We compared our results with those of the heavy baryon (HB) ChPT and found that on average the lattice QCD quark mass dependent data can be better described by
the covariant BChPT, consistent with previous studies. In addition, we found that the baryon pole diagram, which is absent in the HB ChPT,  can
play an important role in certain cases.

Compared with the results of other approaches, our predicted magnetic moments for the anti-triplets are relatively small. The same is true for the
$\Sigma_c^{++}$, $\Sigma_c^+$, and $\Xi_c^{'+}$. On the other hand, our results for $\Sigma^0_c$, $\Xi_c^{'0}$, and $\Omega_c^0$ are relatively large (small in absolute value).  It is not clear how to understand such a pattern at present. We hope that future lattice QCD or experimental studies can
help us gain more insight into these important quantities and better understand the singly charmed baryons.

\section{Acknowledgements}
RXS thanks Jun-Xu Lu and Xiu-Lei Ren for useful discussions.
This work is partly supported by the National Natural Science Foundation of China under Grants No.11522539, No. 11735003, and the fundamental Research Funds for the Central Universities.

\section{Appendix}
The pertinent loop functions, with the PCB terms removed, are given here.
\vspace{8cm}
\begin{figure*}
\begin{eqnarray}
&&H_{B_{\bar{3}}}^{(b)}(\delta_1,m_\phi)=H^{(b)}(m_{\bar{3}},\delta_1,m_\phi),\qquad H_{B_{\bar{3}}}^{(d)}(\delta_1,m_\phi)=H^{(d)}(m_{\bar{3}},\delta_1,m_\phi),\nonumber\\
&&H_{B_{\bar{3}}}^{(c)}(\delta_2,m_\phi)=\left\{
\begin{array}{c}
H_{\delta<m_\phi}^{(c)}(m_{\bar{3}},\delta_2,m_\phi),\qquad (\delta_2<m_\phi)\\
H_{\delta>m_\phi}^{(c)}(m_{\bar{3}},\delta_2,m_\phi),\qquad (\delta_2>m_\phi)
\end{array}
\right.\nonumber\\
&&H_{B_{\bar{3}}}^{(e)}(\delta_2,m_\phi)=\left\{
\begin{array}{c}
H_{\delta<m_\phi}^{(e)}(m_{\bar{3}},\delta_2,m_\phi),\qquad (\delta_2<m_\phi)\\
H_{\delta>m_\phi}^{(e)}(m_{\bar{3}},\delta_2,m_\phi),\qquad (\delta_2>m_\phi)
\end{array}
\right.\nonumber\\
&&H_{B_6}^{(b)}(0,m_\phi)=H^{(b)}(m_6,0,m_\phi),\qquad H_{B_6}^{(b)}(\delta_1,m_\phi)=H^{(b)}(m_6,-\delta_1,m_\phi)\nonumber\\
&&H_{B_6}^{(d)}(0,m_\phi)=H^{(d)}(m_6,0,m_\phi),\qquad H_{B_6}^{(d)}(\delta_1,m_\phi)=H^{(d)}(m_6,-\delta_1,m_\phi)\nonumber\\
&&H_{B_6}^{(c)}(\delta_3,m_\phi)=H_{\delta<m_\phi}^{(c)}(m_6,\delta_3,m_\phi),\qquad (\delta_3<m_\phi)\nonumber\\
&&H_{B_6}^{(e)}(\delta_3,m_\phi)=H_{\delta<m_\phi}^{(e)}(m_6,\delta_3,m_\phi),\qquad (\delta_3<m_\phi)
\end{eqnarray}

\begin{eqnarray}
H^{(b)}(m_B,0,m_\phi)&=&-\frac{1}{4\pi^2}\left[2m_{\phi}^2+\frac{m_{\phi}^2}{m_{B}^2}(2m_{B}^2-m_{\phi}^2)
\log\left(\frac{m_{\phi}^2}{m_{B}^2}\right)\right.
+\left.\frac{2m_{\phi}(m_{\phi}^4-4m_{\phi}^2m_{B}^2+2m_{B}^2)}{m_{B}^2\sqrt{4m_{B}^2-m_{\phi}^2}}
\arccos(\frac{m_{\phi}}{2m_{B}})\right],
\end{eqnarray}

\begin{eqnarray}
H^{(b)}(m_B,\delta,m_\phi)&=&\frac{m_B}{8 \pi ^2}\int_0^1dx\int_0^{1-x}dy\left\{\frac{x^4 m_B^3+\delta  x^3 m_B^2}{x \left(m_B+\delta \right){}^2+(x-1) \left(x m_B^2-m_{\phi }^2\right)}\right.\nonumber\\
&&+\left[\left(4 x^2+4 x-2\right) m_B+\delta  (3 x-1)\right] \log \left(\frac{x \left(m_B+\delta \right){}^2+(x-1) \left(x m_B^2-m_{\phi }^2\right)}{\mu ^2}\right)\nonumber\\
&&\left.-2 \left(2 x^2+2 x-1\right) m_B \log \left(\frac{x^2 m_B^2}{\mu ^2}\right)-x^2 m_B-\delta +4 \delta  x\right\},\qquad(0<\delta<m_\phi)
\end{eqnarray}

\begin{eqnarray}
H^{(d)}(m_B,0,m_\phi)&=&-\frac{1}{4\pi^2}\left[2m_{\phi}^{2}+\frac{m_{\phi}^2}{m_{B}^2}(m_{B}^2-m_{\phi}^2)
\log\left(\frac{m_{\phi}^2}{m_{B}^2}\right)\right.
+\left.\frac{2m_{\phi}^3(m_{\phi}^2-3m_{B}^2)}{m_{B}^{2}\sqrt{4m_{B}^2-m_{\phi}^2}}\arccos(\frac{m_{\phi}}{2m_{B}})\right],
\end{eqnarray}

\begin{eqnarray}
H^{(d)}(m_B,\delta,m_\phi)&=&-\frac{\left(2 m_B+\delta \right){}^2}{16 \pi ^2 m_B^4} \left\{2 \left[m_B^2 \left(m_{\phi }^2-2 \delta ^2\right)+3 \delta  m_B \left(m_{\phi }-\delta \right) \left(\delta +m_{\phi }\right)-\left(m_{\phi }^2-\delta ^2\right){}^2\right] \log \left(\frac{m_{\phi }}{m_B+\delta }\right)\right.\nonumber\\
&&\left.-2 \sqrt{\frac{\left(m_{\phi }-\delta \right) \left(\delta +m_{\phi }\right)}{\left(2 m_B+\delta -m_{\phi }\right) \left(2 m_B+\delta +m_{\phi }\right)}} \left[m_B^2 \left(8 \delta ^2-3 m_{\phi }^2\right)+5 \delta  m_B \left(\delta ^2-m_{\phi }^2\right)+4 \delta  m_B^3\right.\right.\nonumber\\
&&\left.+\left(m_{\phi }^2-\delta ^2\right){}^2\right] \tan ^{-1}\left(\frac{-2 \delta  m_B-2 m_B^2-\delta ^2+m_{\phi }^2}{\sqrt{\left(m_{\phi }-\delta \right) \left(\delta +m_{\phi }\right) \left(2 m_B+\delta -m_{\phi }\right) \left(2 m_B+\delta +m_{\phi }\right)}}\right)\nonumber\\
&&+2 \left[m_B^2 \left(8 \delta ^2-3 m_{\phi }^2\right)+5 \delta  m_B \left(\delta ^2-m_{\phi }^2\right)+4 \delta  m_B^3+\left(m_{\phi }^2-\delta ^2\right){}^2\right]\nonumber\\
&&\cdot\sqrt{\frac{\left(m_{\phi }-\delta \right) \left(\delta +m_{\phi }\right)}{\left(2 m_B+\delta -m_{\phi }\right) \left(2 m_B+\delta +m_{\phi }\right)}}\nonumber\\
&&\cdot\tan ^{-1}\left(\frac{m_{\phi }^2-\delta  \left(2 m_B+\delta \right)}{\sqrt{\left(m_{\phi }-\delta \right) \left(\delta +m_{\phi }\right) \left(2 m_B+\delta -m_{\phi }\right) \left(2 m_B+\delta +m_{\phi }\right)}}\right)\nonumber\\
&&\left.+m_B^2 \left[-2 \delta  m_B+m_B^2+2 \left(m_{\phi }-\delta \right) \left(\delta +m_{\phi }\right)\right]\right\}+\frac{m_B^2}{4\pi^2},
\end{eqnarray}
\end{figure*}

\begin{figure*}
\begin{eqnarray}
&&H_{\delta<m_\phi}^{(c)}(m_B,\delta,m_\phi)\nonumber\\
&=&\frac{1}{864 \pi ^2 m_B^4 \left(\delta +m_B\right){}^2}\left\{-30 \log \left(\frac{\delta +m_B}{m_{\phi }}\right) m_{\phi }^8-6 \left[\left(38 \log \left(\frac{m_{\phi }}{\delta +m_B}\right)+5\right) m_B^2\right.\right.\nonumber\\
&&+5 \left(\tan ^{-1}\left(\frac{\delta ^2+2 m_B \delta -m_{\phi }^2}{\sqrt{\left(\left(\delta +2 m_B\right){}^2-m_{\phi }^2\right) \left(m_{\phi }^2-\delta ^2\right)}}\right)-\tan ^{-1}\left(\frac{\delta ^2+2 m_B \delta +2 m_B^2-m_{\phi }^2}{\sqrt{\left(\left(\delta +2 m_B\right){}^2-m_{\phi }^2\right) \left(m_{\phi }^2-\delta ^2\right)}}\right)\right)\nonumber\\
&&\left.\cdot\sqrt{\left(\left(\delta +2 m_B\right){}^2-m_{\phi }^2\right) \left(m_{\phi }^2-\delta ^2\right)}-4 \delta  \log \left(\frac{\delta +m_B}{m_{\phi }}\right) \left(5 \delta +11 m_B\right)\right] m_{\phi }^6\nonumber\\
&&+3 \left[-4 \log \left(\frac{\delta +m_B}{m_{\phi }}\right) \left(15 \delta ^2+66 m_B \delta +79 m_B^2\right) \delta ^2+2 \left(\tan ^{-1}\left(\frac{\delta ^2+2 m_B \delta -m_{\phi }^2}{\sqrt{\left(\left(\delta +2 m_B\right){}^2-m_{\phi }^2\right) \left(m_{\phi }^2-\delta ^2\right)}}\right)\right.\right.\nonumber\\
&&\left.-\tan ^{-1}\left(\frac{\delta ^2+2 m_B \delta +2 m_B^2-m_{\phi }^2}{\sqrt{\left(\left(\delta +2 m_B\right){}^2-m_{\phi }^2\right) \left(m_{\phi }^2-\delta ^2\right)}}\right)\right) \sqrt{\left(\left(\delta +2 m_B\right){}^2-m_{\phi }^2\right) \left(m_{\phi }^2-\delta ^2\right)}\nonumber\\
&&\cdot\left(15 \delta ^2+34 m_B \delta +28 m_B^2\right)+m_B^2 \left(30 \delta ^2+68 m_B \delta +61 m_B^2+2 \log \left(\frac{m_{\phi }}{\delta +m_B}\right)\right.\nonumber\\
&&\left.\left.\cdot\left(76 \delta ^2+195 m_B \delta +75 m_B^2\right)\right)\right] m_{\phi }^4\nonumber\\
&&+2 \left[30 \log \left(\delta +m_B\right) \left(12 \delta +m_B\right) m_B^5\right.-\left(45 \delta ^4+204 m_B \delta ^3+354 m_B^2 \delta ^2+267 m_B^3 \delta +92 m_B^4\right) m_B^2\nonumber\\
&&+3 \left(\tan ^{-1}\left(\frac{\delta ^2+2 m_B \delta +2 m_B^2-m_{\phi }^2}{\sqrt{\left(\left(\delta +2 m_B\right){}^2-m_{\phi }^2\right) \left(m_{\phi }^2-\delta ^2\right)}}\right)-\tan ^{-1}\left(\frac{\delta ^2+2 m_B \delta -m_{\phi }^2}{\sqrt{\left(\left(\delta +2 m_B\right){}^2-m_{\phi }^2\right)\left(m_{\phi }^2-\delta ^2\right)}}\right)\right)\nonumber\\
&&\sqrt{\left(\left(\delta +2 m_B\right){}^2-m_{\phi }^2\right) \left(m_{\phi }^2-\delta ^2\right)} \left(15 \delta ^4+68 m_B \delta ^3+118 m_B^2 \delta ^2+91 m_B^3 \delta +29 m_B^4\right)\nonumber\\
&&+3 \left(-2 \log \left(\frac{1}{\mu ^2}\right) \left(3 \delta +2 m_B\right) m_B^5-6 \log \left(m_{\phi }\right) \left(22 \delta +3 m_B\right) m_B^5\right.\nonumber\\
&&-\delta ^3 \log \left(\frac{m_{\phi }}{\delta +m_B}\right) \left(38 \delta +195 m_B\right) m_B^2+\delta ^2 \log \left(\frac{\delta +m_B}{m_{\phi }}\right)\nonumber\\
&&\left.\left.\cdot\left(20 \delta ^4+132 m_B \delta ^3+316 m_B^2 \delta ^2+295 m_B^3 \delta +360 m_B^4\right)\right)\right] m_{\phi }^2\nonumber\\
&&\left[-72 \delta  \log \left(m_{\phi }\right) m_B^6 \left(\delta +2 m_B\right)+60 \log \left(\delta +m_B\right) m_B^6 \left(2 \delta +m_B\right) \left(\delta +2 m_B\right)\right.\nonumber\\
&&+6 \log \left(\frac{1}{\mu ^2}\right) m_B^6 \left(\delta +2 m_B\right) \left(4 \delta +5 m_B\right)+30 \delta ^3 \log \left(\frac{m_{\phi }}{\delta +m_B}\right) m_B^4 \left(59 \delta +26 m_B\right)\nonumber\\
&&-6 \delta ^5 \log \left(\frac{\delta +m_B}{m_{\phi }}\right) \left(5 \delta ^3+44 m_B \delta ^2+158 m_B^2 \delta +295 m_B^3\right)\nonumber\\
&&+6 \left(\tan ^{-1}\left(\frac{\delta ^2+2 m_B \delta -m_{\phi }^2}{\sqrt{\left(\left(\delta +2 m_B\right){}^2-m_{\phi }^2\right) \left(m_{\phi }^2-\delta ^2\right)}}\right)-\tan ^{-1}\left(\frac{\delta ^2+2 m_B \delta +2 m_B^2-m_{\phi }^2}{\sqrt{\left(\left(\delta +2 m_B\right){}^2-m_{\phi }^2\right) \left(m_{\phi }^2-\delta ^2\right)}}\right)\right)\nonumber\\
&&\cdot\sqrt{\left(\left(\delta +2 m_B\right){}^2-m_{\phi }^2\right) \left(m_{\phi }^2-\delta ^2\right)} \left(\delta +2 m_B\right){}^2 \left(5 \delta ^4+14 m_B \delta ^3+14 m_B^2 \delta ^2+3 m_B^3 \delta -3 m_B^4\right)\nonumber\\
&&\left.\left.+m_B^2 \left(30 \delta ^6+204 m_B \delta ^5+525 m_B^2 \delta ^4+618 m_B^3 \delta ^3+250 m_B^4 \delta ^2-110 m_B^5 \delta -86 m_B^6\right)\right]\right\}\nonumber\\
&&-\frac{m_B^2 \left(30 \log \left(\frac{m_B^2}{\mu ^2}\right)-43\right)}{432 \pi ^2},
\end{eqnarray}
\vspace{3cm}
\end{figure*}

\begin{figure*}
\begin{eqnarray}
&&H_{\delta>m_\phi}^{(c)}(m_B,\delta,m_\phi)\nonumber\\
&=&\frac{1}{864 \pi ^2 m_B^4 \left(\delta +m_B\right){}^2}\left\{-30 \log \left(\frac{\delta +m_B}{m_{\phi }}\right) m_{\phi }^8-6 \left[\left(38 \log \left(\frac{m_{\phi }}{\delta +m_B}\right)+5\right) m_B^2\right.\right.\nonumber\\
&&+5 \coth ^{-1}\left(\frac{m_{\phi }^2+\delta  \left(\delta +2 m_B\right)}{\sqrt{\left(\delta ^2-m_{\phi }^2\right) \left(\left(\delta +2 m_B\right){}^2-m_{\phi }^2\right)}}\right) \sqrt{\left(\delta ^2-m_{\phi }^2\right) \left(\left(\delta +2 m_B\right){}^2-m_{\phi }^2\right)}\nonumber\\
&&\left.-4 \delta  \log \left(\frac{\delta +m_B}{m_{\phi }}\right) \left(5 \delta +11 m_B\right)\right] m_{\phi }^6\nonumber\\
&&+3 \left[-4 \log \left(\frac{\delta +m_B}{m_{\phi }}\right) \left(15 \delta ^2+66 m_B \delta +79 m_B^2\right) \delta ^2+2 \coth ^{-1}\left(\frac{m_{\phi }^2+\delta  \left(\delta +2 m_B\right)}{\sqrt{\left(\delta ^2-m_{\phi }^2\right) \left(\left(\delta +2 m_B\right){}^2-m_{\phi }^2\right)}}\right)\right.\nonumber\\
&&\cdot\sqrt{\left(\delta ^2-m_{\phi }^2\right) \left(\left(\delta +2 m_B\right){}^2-m_{\phi }^2\right)} \left(15 \delta ^2+34 m_B \delta +28 m_B^2\right)+m_B^2 \left(30 \delta ^2+68 m_B \delta +61 m_B^2\right.\nonumber\\
&&\left.\left.+2 \log \left(\frac{m_{\phi }}{\delta +m_B}\right) \left(76 \delta ^2+195 m_B \delta +75 m_B^2\right)\right)\right] m_{\phi }^4\nonumber\\
&&+2 \left[30 \log \left(\delta +m_B\right) m_B^6\right.-\left(45 \delta ^4+204 m_B \delta ^3+354 m_B^2 \delta ^2+267 m_B^3 \delta +92 m_B^4\right) m_B^2\nonumber\\
&&-3 \coth ^{-1}\left(\frac{m_{\phi }^2+\delta  \left(\delta +2 m_B\right)}{\sqrt{\left(\delta ^2-m_{\phi }^2\right) \left(\left(\delta +2 m_B\right){}^2-m_{\phi }^2\right)}}\right) \left(15 \delta ^4+68 m_B \delta ^3+118 m_B^2 \delta ^2+91 m_B^3 \delta +29 m_B^4\right)\nonumber\\
&&\cdot\sqrt{\left(\delta ^2-m_{\phi }^2\right) \left(\left(\delta +2 m_B\right){}^2-m_{\phi }^2\right)}+3 \left(20 \log \left(\frac{\delta +m_B}{m_{\phi }}\right) \delta ^6+132 \log \left(\frac{\delta +m_B}{m_{\phi }}\right) m_B \delta ^5\right.\nonumber\\
&&+316 \log \left(\frac{\delta +m_B}{m_{\phi }}\right) m_B^2 \delta ^4+295 \log \left(\frac{\delta +m_B}{m_{\phi }}\right) m_B^3 \delta ^3-\log \left(\frac{m_{\phi }}{\delta +m_B}\right) m_B^2 \left(38 \delta +195 m_B\right) \delta ^3\nonumber\\
&&+360 \log \left(\frac{\delta +m_B}{m_{\phi }}\right) m_B^4 \delta ^2-6 \log \left(\frac{\left(\delta +m_B\right){}^2}{\mu ^2}\right) m_B^5 \delta +132 \log \left(\frac{\delta +m_B}{m_{\phi }}\right) m_B^5 \delta\nonumber\\
&&\left.\left.-4 \log \left(\frac{1}{\mu ^2}\right) m_B^6-18 \log \left(m_{\phi }\right) m_B^6\right)\right] m_{\phi }^2\nonumber\\
&&+60 \delta  \log \left(\delta +m_B\right) m_B^6 \left(2 \delta +5 m_B\right)+6 \coth ^{-1}\left(\frac{m_{\phi }^2+\delta  \left(\delta +2 m_B\right)}{\sqrt{\left(\delta ^2-m_{\phi }^2\right) \left(\left(\delta +2 m_B\right){}^2-m_{\phi }^2\right)}}\right) \nonumber\\
&&\sqrt{\left(\delta ^2-m_{\phi }^2\right) \left(\left(\delta +2 m_B\right){}^2-m_{\phi }^2\right)} \left(\delta +2 m_B\right){}^2 \left(5 \delta ^4+14 m_B \delta ^3\right.+14 m_B^2 \delta ^2+3 m_B^3 \delta -3 m_B^4)\nonumber\\
&&+m_B^2 \left(30 \delta ^6+204 m_B \delta ^5+525 m_B^2 \delta ^4+618 m_B^3 \delta ^3+250 m_B^4 \delta ^2-110 m_B^5 \delta -86 m_B^6\right)\nonumber\\
&&+6 \left[10 \log \left(\frac{\left(\delta +m_B\right){}^2}{\mu ^2}\right) m_B^8
-12 \delta  \log \left(m_{\phi }\right) \left(\delta +2 m_B\right) m_B^6+\delta  \log \left(\frac{1}{\mu ^2}\right) \left(4 \delta +13 m_B\right) m_B^6\right.\nonumber\\
&&\left.\left.+5 \delta ^3 \log \left(\frac{m_{\phi }}{\delta +m_B}\right) \left(59 \delta +26 m_B\right) m_B^4-\delta ^5 \log \left(\frac{\delta +m_B}{m_{\phi }}\right) \left(5 \delta ^3+44 m_B \delta ^2+158 m_B^2 \delta +295 m_B^3\right)\right]\right\}\nonumber\\
&&-\frac{m_B^2 \left(30 \log \left(\frac{m_B^2}{\mu ^2}\right)-43\right)}{432 \pi ^2},
\end{eqnarray}
\vspace{6cm}
\end{figure*}

\begin{figure*}
\begin{eqnarray}
H_{\delta<m_\phi}^{(e)}(m_B,\delta,m_\phi)&=&\frac{\left(80 \log \left(\frac{m_B^2}{\mu ^2}\right)-3\right) m_B^2}{432 \pi ^2}+\frac{1}{432 \pi ^2 \left(\delta +m_B\right){}^4 m_B^4}\left\{\left[-2 \log \left(\frac{\delta +m_B}{m_{\phi }}\right) \left(9 \delta ^2+22 m_B \delta +16 m_B^2\right)\right] m_{\phi }^8\right.\nonumber\\
&&+2 \left[-\left(9 \left(28 \log \left(\frac{m_{\phi }}{\delta +m_B}\right)+1\right) \delta ^2+22 m_B \delta +16 m_B^2\right) m_B^2\right.\nonumber\\
&&+\left(\tan ^{-1}\left(\frac{\delta ^2+2 m_B \delta +2 m_B^2-m_{\phi }^2}{\sqrt{\left(\left(\delta +2 m_B\right){}^2-m_{\phi }^2\right) \left(m_{\phi }^2-\delta ^2\right)}}\right)-\tan ^{-1}\left(\frac{\delta ^2+2 m_B \delta -m_{\phi }^2}{\sqrt{\left(\left(\delta +2 m_B\right){}^2-m_{\phi }^2\right) \left(m_{\phi }^2-\delta ^2\right)}}\right)\right)\nonumber\\
&&\cdot\sqrt{\left(\left(\delta +2 m_B\right){}^2-m_{\phi }^2\right) \left(m_{\phi }^2-\delta ^2\right)} \left(9 \delta ^2+22 m_B \delta +16 m_B^2\right)\nonumber\\
&&\left.+4 \log \left(\frac{\delta +m_B}{m_{\phi }}\right) \left(9 \delta ^4+40 m_B \delta ^3+8 m_B^2 \delta ^2+62 m_B^3 \delta +25 m_B^4\right)\right] m_{\phi }^6\nonumber\\
&&+\left[16 \delta ^2 \log \left(\frac{m_{\phi }}{\delta +m_B}\right) \left(63 \delta ^2+137 m_B \delta +135 m_B^2\right) m_B^2\right.\nonumber\\
&&+\left(54 \delta ^4+240 m_B \delta ^3+421 m_B^2 \delta ^2+330 m_B^3 \delta +84 m_B^4\right) m_B^2\nonumber\\
&&+2 \left(\tan ^{-1}\left(\frac{\delta ^2+2 m_B \delta -m_{\phi }^2}{\sqrt{\left(\left(\delta +2 m_B\right){}^2-m_{\phi }^2\right) \left(m_{\phi }^2-\delta ^2\right)}}\right)-\tan ^{-1}\left(\frac{\delta ^2+2 m_B \delta +2 m_B^2-m_{\phi }^2}{\sqrt{\left(\left(\delta +2 m_B\right){}^2-m_{\phi }^2\right) \left(m_{\phi }^2-\delta ^2\right)}}\right)\right)\nonumber\\
&&\sqrt{\left(\left(\delta +2 m_B\right){}^2-m_{\phi }^2\right) \left(m_{\phi }^2-\delta ^2\right)} \left(27 \delta ^4+120 m_B \delta ^3+206 m_B^2 \delta ^2+172 m_B^3 \delta +68 m_B^4\right)\nonumber\\
&&-4 \left(-204 \delta  \log \left(m_{\phi }\right) m_B^5+60 \log \left(\delta +m_B\right) \left(4 \delta +m_B\right) m_B^5+6 \log \left(\frac{1}{\mu ^2}\right) \left(3 \delta +5 m_B\right) m_B^5\right.\nonumber\\
&&\left.\left.+\delta ^3 \log \left(\frac{\delta +m_B}{m_{\phi }}\right) \left(27 \delta ^3+174 m_B \delta ^2+216 m_B^2 \delta +124 m_B^3\right)\right)\right] m_{\phi }^4\nonumber\\
&&-2 \left[20 \log \left(\delta +m_B\right) \left(-85 \delta ^3-27 m_B \delta ^2+2 m_B^2 \delta +2 m_B^3\right) m_B^5\right.\nonumber\\
&&+\left(27 \delta ^6+174 m_B \delta ^5+454 m_B^2 \delta ^4+612 m_B^3 \delta ^3+422 m_B^4 \delta ^2+110 m_B^5 \delta -6 m_B^6\right) m_B^2\nonumber\\
&&+\left(\tan ^{-1}\left(\frac{\delta ^2+2 m_B \delta -m_{\phi }^2}{\sqrt{\left(\left(\delta +2 m_B\right){}^2-m_{\phi }^2\right) \left(m_{\phi }^2-\delta ^2\right)}}\right)-\tan ^{-1}\left(\frac{\delta ^2+2 m_B \delta +2 m_B^2-m_{\phi }^2}{\sqrt{\left(\left(\delta +2 m_B\right){}^2-m_{\phi }^2\right) \left(m_{\phi }^2-\delta ^2\right)}}\right)\right)\nonumber\\
&&\cdot\sqrt{\left(\left(\delta +2 m_B\right){}^2-m_{\phi }^2\right) \left(m_{\phi }^2-\delta ^2\right)} \left(\delta +2 m_B\right)\left(27 \delta ^5+120 m_B \delta ^4+214 m_B^2 \delta ^3+172 m_B^3 \delta ^2+50 m_B^4 \delta +8 m_B^5\right)\nonumber\\
&&+4 \left(\log \left(\frac{m_{\phi }}{\delta +m_B}\right) m_B^2 \left(63 \delta +274 m_B\right) \delta ^5-\log \left(\frac{\delta +m_B}{m_{\phi }}\right) \left(9 \delta ^4+76 m_B \delta ^3+208 m_B^2 \delta ^2+257 m_B^3 \delta +620 m_B^4\right) \delta ^4\right.\nonumber\\
&&\left.\left.+\log \left(\frac{1}{\mu ^2}\right) m_B^5 \left(9 \delta ^3+36 m_B \delta ^2+41 m_B^2 \delta +11 m_B^3\right)+\log \left(m_{\phi }\right) m_B^5 \left(443 \delta ^3+207 m_B \delta ^2+72 m_B^2 \delta +12 m_B^3\right)\right)\right] m_{\phi }^2\nonumber\\
&&+\left[3180 \delta ^6 \log \left(\frac{m_{\phi }}{\delta +m_B}\right) m_B^4+32 \delta ^2 \log \left(m_{\phi }\right) m_B^5 \left(103 \delta ^3+73 m_B \delta ^2+32 m_B^2 \delta +6 m_B^3\right)\right.\nonumber\\
&&-2 \delta ^7 \log \left(\frac{\delta +m_B}{m_{\phi }}\right) \left(9 \delta ^3+94 m_B \delta ^2+416 m_B^2 \delta +1028 m_B^3\right)\nonumber\\
&&+2 \delta  \left(\tan ^{-1}\left(\frac{\delta ^2+2 m_B \delta -m_{\phi }^2}{\sqrt{\left(\left(\delta +2 m_B\right){}^2-m_{\phi }^2\right) \left(m_{\phi }^2-\delta ^2\right)}}\right)-\tan ^{-1}\left(\frac{\delta ^2+2 m_B \delta +2 m_B^2-m_{\phi }^2}{\sqrt{\left(\left(\delta +2 m_B\right){}^2-m_{\phi }^2\right) \left(m_{\phi }^2-\delta ^2\right)}}\right)\right)\nonumber\\
&&\sqrt{\left(\left(\delta +2 m_B\right){}^2-m_{\phi }^2\right) \left(m_{\phi }^2-\delta ^2\right)} \left(\delta +2 m_B\right){}^3 \left(9 \delta ^4+22 m_B \delta ^3+24 m_B^2 \delta ^2+20 m_B^3 \delta +6 m_B^4\right)\nonumber\\
&&-2 \log \left(\frac{1}{\mu ^2}\right) m_B^5 \left(\delta +2 m_B\right) \left(36 \delta ^4+144 m_B \delta ^3+196 m_B^2 \delta ^2+105 m_B^3 \delta +20 m_B^4\right)\nonumber\\
&&-20 \log \left(\delta +m_B\right) m_B^5 \left(172 \delta ^5+160 m_B \delta ^4+148 m_B^2 \delta ^3+109 m_B^3 \delta ^2+46 m_B^4 \delta +8 m_B^5\right)\nonumber\\
&&\left.\left.+m_B^2 \left(18 \delta ^8+152 m_B \delta ^7+519 m_B^2 \delta ^6+918 m_B^3 \delta ^5+954 m_B^4 \delta ^4+688 m_B^5 \delta ^3+382 m_B^6 \delta ^2+116 m_B^7 \delta +3 m_B^8\right)\right]\right\},\nonumber\\
 \end{eqnarray}
 \end{figure*}

 \begin{figure*}
 \begin{eqnarray}
 H_{\delta>m_\phi}^{(e)}(m_B,\delta,m_\phi)&=&\frac{\left(80 \log \left(\frac{m_B^2}{\mu ^2}\right)-3\right) m_B^2}{432 \pi ^2}+\frac{1}{432 \pi ^2 \left(\delta +m_B\right){}^4 m_B^4}\left\{\left[-2 \log \left(\frac{\delta +m_B}{m_{\phi }}\right) \left(9 \delta ^2+22 m_B \delta+16 m_B^2\right)\right] m_{\phi }^8\right.\nonumber\\
 &&+2 \left[-4 \log \left(\frac{m_{\phi }}{\delta +m_B}\right) \left(9 \delta +40 m_B\right) \delta ^3-\coth ^{-1}\left(\frac{\delta ^2+2 m_B \delta +m_{\phi }^2}{\sqrt{\left(\delta ^2-m_{\phi }^2\right) \left(\left(\delta +2 m_B\right){}^2-m_{\phi }^2\right)}}\right) \left(9 \delta ^2+22 m_B \delta +16 m_B^2\right)\right.\nonumber\\
 &&\cdot\sqrt{\left(\delta ^2-m_{\phi }^2\right) \left(\left(\delta +2 m_B\right){}^2-m_{\phi }^2\right)}\nonumber\\
 &&\left.+m_B^2 \left(-9 \delta ^2-22 m_B \delta -16 m_B^2+4 \log \left(\frac{\delta +m_B}{m_{\phi }}\right) \left(71 \delta ^2+62 m_B \delta +25 m_B^2\right)\right)\right] m_{\phi }^6\nonumber\\
 &&+\left[\left(54 \delta ^4+240 m_B \delta ^3+421 m_B^2 \delta ^2+16 \log \left(\frac{m_{\phi }}{\delta +m_B}\right) \left(117 \delta ^2+137 m_B \delta +135 m_B^2\right) \delta ^2\right.\right.\nonumber\\
 &&\left.+330 m_B^3 \delta +84 m_B^4-24 \log \left(\frac{m_{\phi }^2}{\mu ^2}\right) m_B^3 \left(3 \delta +5 m_B\right)\right) m_B^2\nonumber\\
 &&+2 \coth ^{-1}\left(\frac{\delta ^2+2 m_B \delta +m_{\phi }^2}{\sqrt{\left(\delta ^2-m_{\phi }^2\right) \left(\left(\delta +2 m_B\right){}^2-m_{\phi }^2\right)}}\right) \sqrt{\left(\delta ^2-m_{\phi }^2\right) \left(\left(\delta +2 m_B\right){}^2-m_{\phi }^2\right)}\nonumber\\
 &&\cdot\left(27 \delta ^4+120 m_B \delta ^3+206 m_B^2 \delta ^2+172 m_B^3 \delta +68 m_B^4\right)\nonumber\\
 &&\left.-4 \log \left(\frac{\delta +m_B}{m_{\phi }}\right) \left(27 \delta ^6+174 m_B \delta ^5+124 m_B^3 \delta ^3+240 m_B^5 \delta +60 m_B^6\right)\right] m_{\phi }^4\nonumber\\
 &&-2\left[\left(27 \delta ^6+174 m_B \delta ^5+454 m_B^2 \delta ^4+612 m_B^3 \delta ^3+422 m_B^4 \delta ^2+110 m_B^5 \delta -6 m_B^6\right) m_B^2\right.\nonumber\\
 &&+\coth ^{-1}\left(\frac{\delta ^2+2 m_B \delta +m_{\phi }^2}{\sqrt{\left(\delta ^2-m_{\phi }^2\right)\left(\left(\delta +2 m_B\right){}^2-m_{\phi }^2\right)}}\right) \sqrt{\left(\delta ^2-m_{\phi }^2\right) \left(\left(\delta +2 m_B\right){}^2-m_{\phi }^2\right)} \left(\delta +2 m_B\right)\nonumber\\
 &&\cdot \left(27 \delta ^5+120 m_B \delta ^4+214 m_B^2 \delta ^3+172 m_B^3 \delta ^2+50 m_B^4 \delta +8 m_B^5\right)\nonumber\\
 &&+4 \log \left(\frac{1}{\mu ^2}\right) \left(11 m_B^8+41 \delta  m_B^7+9 \delta ^3 m_B^5\right)+20 \log \left(\delta +m_B\right) \left(2 m_B^8+2 \delta  m_B^7-85 \delta ^3 m_B^5\right)\nonumber\\
 &&+4 \left(-\log \left(\frac{\delta +m_B}{m_{\phi }}\right) m_B^2 \left(271 \delta ^2+257 m_B \delta +620 m_B^2\right) \delta ^4\right.\nonumber\\
 &&+\left(36 \log \left(\frac{m_{\phi }^2}{\mu ^2}\right) m_B^6+\log \left(\frac{m_{\phi }}{\delta +m_B}\right) \left(9 \delta ^6+76 m_B \delta ^5+274 m_B^3 \delta ^3+135 m_B^6\right)\right) \delta ^2\nonumber\\
 &&\left.\left.+\log \left(m_{\phi }\right) \left(12 m_B^8+72 \delta  m_B^7+443 \delta ^3 m_B^5\right)\right)\right]m_{\phi }^2\nonumber\\
 &&+\left[-20 \left(\log \left(\frac{1}{\mu ^2}\right)+2 \log \left(\delta +m_B\right)\right) \left(23 \delta +4 m_B\right) m_B^9\right.\nonumber\\
 &&-2 \delta ^2 \log \left(\frac{m_{\phi }^2}{\mu ^2}\right) \left(36 \delta ^3+216 m_B \delta ^2+484 m_B^2 \delta +497 m_B^3\right) m_B^5+4 \delta ^6 \log \left(\frac{m_{\phi }}{\delta +m_B}\right) \left(208 \delta ^2+795 m_B^2\right) m_B^2\nonumber\\
 &&+\left(18 \delta ^8+152 m_B \delta ^7+519 m_B^2 \delta ^6+918 m_B^3 \delta ^5+954 m_B^4 \delta ^4+688 m_B^5 \delta ^3+382 m_B^6 \delta ^2+116 m_B^7 \delta +3 m_B^8\right) m_B^2\nonumber\\
 &&+2 \delta  \coth ^{-1}\left(\frac{\delta ^2+2 m_B \delta +m_{\phi }^2}{\sqrt{\left(\delta ^2-m_{\phi }^2\right) \left(\left(\delta +2 m_B\right){}^2-m_{\phi }^2\right)}}\right) \sqrt{\left(\delta ^2-m_{\phi }^2\right) \left(\left(\delta +2 m_B\right){}^2-m_{\phi }^2\right)}\nonumber\\
 &&\cdot\left(\delta +2 m_B\right){}^3 \left(9 \delta ^4+22 m_B \delta ^3+24 m_B^2 \delta ^2+20 m_B^3 \delta +6 m_B^4\right)\nonumber\\
 &&\left.\left.-2 \delta ^2 \log \left(\frac{\delta +m_B}{m_{\phi }}\right) \left(9 \delta ^8+94 m_B \delta ^7+1028 m_B^3 \delta ^5+1720 m_B^5 \delta ^3+1600 m_B^6 \delta ^2+1480 m_B^7 \delta +1090 m_B^8\right)\right]\right\}.\nonumber\\
 \end{eqnarray}
\end{figure*}

\end{document}